\documentclass[twocolumn]{autart}    

\usepackage[dvips]{epsfig}    

\usepackage{graphicx}%
\usepackage{multirow}%
\usepackage{amsmath,amssymb,amsfonts}%
\usepackage{mathrsfs}%
\usepackage[title]{appendix}%
\usepackage{xcolor}%
\usepackage{textcomp}%
\usepackage{manyfoot}%
\usepackage{booktabs}%
\usepackage{algorithm}%
\usepackage{algorithmicx}%
\usepackage{algpseudocode}%
\usepackage{listings}%
\usepackage{subcaption}
\usepackage{mathbbol}
\usepackage{balance}

\begin{document}

\begin{frontmatter}

\title{Bio-inspired density control of multi-agent swarms via leader-follower plasticity} 


\author[ssm]{Gian Carlo Maffettone}\ead{giancarlo.maffettone@unina.it},%
\author[nyit]{ Alain Boldini}\ead{aboldini@nyit.edu}, \author[unina,ssm]{Mario di Bernardo}\ead{mario.dibernardo@unina.it}, \author[nyu]{Maurizio Porfiri}\ead{mporfiri@nyu.edu} 

\address[ssm]{Modeling and Engineering Risk and Complexity, Scuola Superiore Meridionale, Naples, Italy}  
\address[nyit]{Department of Mechanical Engineering, New York Institute of Technology, Old Westbury, NY, USA}  

\address[unina]{Department of Electrical Engineering and Information Technology, University of Naples Federico II, Naples, Italy} 

\address[nyu]{Department of Mechanical and Aerospace Engineering, Department of Biomedical Engineering, Center for Urban Science and Progress, Tandon School of Engineering, New York University, Brooklyn, NY, USA} 


\begin{abstract}              
The design of control systems for the spatial self-organization of mobile agents is an open challenge across several engineering domains, including swarm robotics and synthetic biology. Here, we propose a bio-inspired leader-follower solution, which is aware of energy constraints of mobile agents and is apt to deal with large swarms. Akin to many natural systems, control objectives are formulated for the entire collective, and leaders and followers are allowed to plastically switch their role in time. We frame a density control problem, modeling the agents' population via a system of nonlinear partial differential equations. This approach allows for a compact description that inherently avoids the curse of dimensionality and improves analytical tractability. We derive analytical guarantees for the existence of desired steady-state solutions and their {global} stability for one-dimensional and higher-dimensional problems.
We numerically validate our control methodology, offering  support to the effectiveness, robustness, and versatility of our proposed bio-inspired control strategy.
\end{abstract}

\end{frontmatter}

\section{Introduction}
{Across applications, from swarm robotics \cite{dorigo2021swarm} to 
search-and-rescue operations \cite{wong2005multi}, leader-follower strategies \cite{hu2010distributed} are used for the spatial displacement of large swarms, in which controller agents (leaders) are to induce a desired spatial behavior of another group of agents (followers). While often informed by biological systems like fish \cite{mugica2022scale,zienkiewicz2015leadership} and birds \cite{pettit2015speed}, engineered systems fail to capture two key aspects of their biological counterparts.

First, leaders and followers maintain fixed hierarchies \cite{maffettone2024leader} missing behavioral plasticity, the reversible mechanism behind role changes in biological systems due to internal or environmental stimuli \cite{price2003role}. In fish schools, plasticity arises when followers access novel information about predators or food \cite{couzin2005effective}; likewise, in migrating birds, leaders step down from frontal positions in the flock to recover energy \cite{voelkl2015matching}. 
Second, engineered leaders typically act as fixed open-loop signals, unreceptive to the unfolding collective dynamics \cite{wang2002pinning}, whereas biological leaders adapt based on collective needs critical to survival, such as in foraging and predator avoidance \cite{camazine2020self}. Capturing these aspects would benefit applications in spatial organization of swarms, granting flexibility to address multiple tasks at lower control efforts and energy costs \cite{hunt2020phenotypic}.
}

Mathematically, the control of large swarms {poses} several challenges: ($i$) the state space grows exponentially with the number of agents; ($ii$) communication graphs may be time-varying and constrained; and ($iii$) individual agent dynamics are { non-locally} coupled. Continuum approaches using partial differential equations (PDEs) offer a promising avenue to circumvent these difficulties by modeling agent densities rather than individual states, achieving dimension reduction preserving essential collective dynamics \cite{maffettone2022continuification,Elamvazhuthi2020,Kolpas2007}.
Mean-field models of interacting populations of agents have gained substantial momentum {in } the last decade \cite{ascione2023mean,albi2014boltzmann,lama2025interpretable}, as they offer compact formulations to increase computational and analytical tractability. Bio-inspired switching mechanisms between leaders and followers have been explored in \cite{albi2024kinetic,albi2019leader,Loy2021boltzmann,bernardi2021leadership,cristiani2025kinetic} for modeling purposes, { lacking the presentation of} feedback control actions to induce desired collective behaviors. A mean-field optimal control problem with transient leadership has been formulated in \cite{albi2022mean}, but, due to the absence of closed form feedback solutions, it can only be numerically approximated, yielding limited insight {informing} control design in realistic settings.
Related work on leader selection \cite{franchi2016online,li2016affection} and switching mechanisms \cite{park2020synchronization} exists, but these approaches either maintain fixed population assignments or focus on synchronization rather than spatial density control.

{Here, we develop a continuum model for a large population of behaviorally plastic leaders and followers solving a density control problem.} We draw insight from the literature on reacting mixtures \cite{mills1966incompressible}, used to describe blood flows \cite{massoudi2012modeling} and tumors growth \cite{faghihi2020coupled}. In particular, we model plasticity as a chemical reaction taking place between two fluids, associated with the continuum description of leaders and followers. We assume that plasticity is not common to all agents, where some of them may not be allowed switching their role. Whether or not an agent is a leader or a follower is not distinguishable by the rest of the group. We derive conditions for the existence and stability of the solution, in terms of key parameters that can be part of engineering design. Our approach provides explicit feedback control laws that guarantee convergence through adaptive leader selection.

The rest of the paper is organized as follows. In Sec. \ref{secch6:prob_stat}, we present the problem statement for a one-dimensional (1D) scenario. The control strategy is formulated in Sec. \ref{secch6:control_des} and { validated} in Sec. \ref{sec:num_valid_1d_React}. An agent-based model { supporting} our continuum approach is presented in Sec. \ref{sec:abm}. The extension to higher dimensions and {its} validation are expounded in Sec. \ref{sec_ch6:higher_dim}. Section \ref{sec:conclusions} concludes the manuscript.

\section{Problem Statement}\label{secch6:prob_stat}


Our control objective is to steer the spatial distribution {of a large-scale swarm} towards a desired configuration through coordinated leader actions and adaptive role assignment. We adopt a continuum approach  consisting of coupled convection-{reaction}-diffusion equations on the unit circle $\mathcal{S}=[-\pi, \pi]$(for the sake of brevity, throughout the paper, explicit dependencies on spatial coordinate $x$ and time $t$ are omitted when unambiguous.)
\begin{subequations}\label{eq:behavioral_plasticity_model}
\begin{align}
    \rho^L_t + \left(\rho^L u \right)_x + \left[\rho^L (f*\rho)\right]_x &= D\rho^L_{xx} +  q,\label{eq:ch6_leaders}\\
    \rho^F_t + \left[\rho^F (f*\rho)\right]_x &= D\rho^F_{xx} - q,\label{eq:ch6_followers}\\
    \eta^F_t + \left[\eta^F (f*\rho)\right]_x &= D\eta^F_{xx},\label{eq:ch6_non_react_followers}
\end{align}
\end{subequations}
where $(\cdot)_t$ and $(\cdot)_x$ denote partial derivatives. {The densities $\rho^L$, $\rho^F$, $\eta^F:\mathcal{S}\times\mathbb{R}_{\geq0}\to \mathbb{R}_{\geq0}$ are associated with three agents' subsets: \textit{leaders} controlled through the periodic velocity field $u:\mathcal{S}\times\mathbb{R}_{\geq0}\to \mathbb{R}$ that may become followers; \textit{plastic followers} that may become leaders; and \textit{non-plastic followers} that cannot be engaged in roles' switching. Plasticity between leaders and followers take place through the reaction term $q:\mathcal{S}\times\mathbb{R}_{\geq0}\to \mathbb{R}$; inter- and intra-population interactions are modeled through a cross-convectional non-local term involving the interaction kernel $f$ (soft-core and odd) and the swarm density $\rho= \rho^L + \rho^F +\eta^F$; {specifically, $(f*\rho)=\int_\mathcal{S
} f(\{x,y\})\rho(y,t)\,\mathrm{d}y$, where $\{x,y\}$ is the relative position between $x$ and $y$ wrapped on $\mathcal{S}$ (due to periodicity, it is a circular convolution)}; agent-level stochasticity is captured by diffusion weighted by the coefficient $D>0$.

Summing eqs. \eqref{eq:behavioral_plasticity_model} yields the total population density dynamics 
\begin{equation}\label{eq:ch6_overall_density}
    \rho_t + \left(\rho^L u \right)_x + \left[\rho (f*\rho)\right]_x = D\rho_{xx},
\end{equation}
which is independent of $q$. Periodic boundary conditions,
\begin{subequations}\label{eq:BC}
    \begin{align}
        \rho^i(-\pi, t) &= \rho^i(\pi, t), \;\forall t\in\mathbb{R}_{\geq 0}, \,i=L, F\\
        \eta^F(-\pi, t) &= \eta^F(\pi, t), \;\forall t\in\mathbb{R}_{\geq 0},
    \end{align}
\end{subequations}
ensure mass conservation, that is $\left(\int_\mathcal{S}\rho\,\mathrm{d}x\right)_t = 0$ (exploiting periodicity of the functions).}
{Note that \eqref{eq:ch6_non_react_followers} satisfies mass conservation on its own, but \eqref{eq:ch6_leaders} and \eqref{eq:ch6_followers} do not, due to the reaction $q$}. If $\int_{\mathcal{S}} q\,\mathrm{d}x\,{=0}$, there is no net mass transfer between leaders and followers. 
Eqs. ~\eqref{eq:behavioral_plasticity_model} are also complemented by the initial conditions
    \begin{align}\label{eq:IC}
    \rho^i(x, 0) = \rho^i_0(x),\,i=L, F,\quad
    \eta^F(x, 0) = \eta^F_0(x).
\end{align}

{Preserving} generality, we normalize the total mass to one
\begin{equation}\label{eq:mass_sum}
    \int_\mathcal{S}\rho(x, t)\,\mathrm{d}x = M^L(t) + M^F(t) + \Phi^F = 1,
\end{equation}
where $M^L$ { and $M^F$ are} the leaders' { and followers'} mass, and $\Phi^F$ is the constant mass of non-plastic followers. {The fraction of the population
allowed to switch role}, $p$, is
\begin{align}\label{eqch6:plasticity}
    p = 1-\Phi^F.
\end{align}

We consider the density control problem of choosing $p$, $u$, and $q$ in \eqref{eq:behavioral_plasticity_model} so that the population density asymptotically converges almost everywhere (a.e.) towards a desired time-invariant density profile $\bar{\rho}:\mathcal{S}\to\mathbb{R}_{\geq 0 }$,
\begin{equation}\label{eq:prob_stat}
    \lim_{t\to\infty} \Vert \bar{\rho}(\cdot) - \rho(\cdot, t)\Vert_2 = 0,
\end{equation}
where $\Vert\cdot\Vert_2$ denotes the $\mathcal{L}^2$ norm over $\mathcal{S}$. We note that the control problem pertains to the density of the entire collective, comprising leaders and followers.
As an additional control specification capturing energy costs, we tune the steady-state leaders-to-followers mass ratio to a desired value $\hat{r}$, namely,
\begin{equation}\label{eq:mass_ratio_requirement}
    \lim_{t\to\infty} M^L(t)/M^F(t) = \hat{r}.
\end{equation}

\section{Bio-inspired Control }\label{secch6:control_des}
\subsection{Design of the leaders' velocity field}
We define the error function for the the collective,
\begin{align}\label{eqch6:e}
    e(x, t) = \bar{\rho}(x) - \rho(x, t).
\end{align}
Using \eqref{eq:ch6_overall_density}, the error dynamics reads
\begin{equation}\label{eqch6:err_dyn}
    e_t = \left(\rho^L u\right)_x +\left[\rho(f*\rho)\right]_x - D\rho_{xx},
\end{equation}
with periodic boundary conditions and initial conditions that can be derived from \eqref{eq:BC} and \eqref{eq:IC}.

\begin{thm}[Global exponential convergence]\label{thch6:err_convergence}%
Assume $\rho_L > 0$ for any $x \in \mathcal{S}$ and $t \in \mathbb{R}_{\geq 0}$. Choosing the control input as
\begin{equation}
\label{eqch6:rho_u}
u = \frac{1}{\rho^L}\left[-K\int e\,dx - \rho(f*\rho) + D\rho_x\right],
\end{equation}
where $K>0$, the error converges to 0 globally and exponentially, that is,
\begin{align}\label{eqch6:err_sol}
    e(x,t) = e(x,0)\exp(-Kt).
\end{align}
\end{thm}
\begin{pf}
    Substituting \eqref{eqch6:rho_u} into \eqref{eqch6:err_dyn}, we obtain
    \begin{align}\label{eq:err_lin}
        e_t(x, t) = - K e(x, t),
    \end{align}
    which is linear and does not involve spatial derivatives. Its analytical solution yields  \eqref{eqch6:err_sol}.\qed
\end{pf}

\begin{rem}
    The control input $u$ is well-defined only for $\rho^L>0$. 
    Note that $u$ can be shown to be periodic (see Corollary 1 in \cite{maffettone2024leader}).
\end{rem}
\begin{rem}\label{remch6:rho}
    {From} \eqref{eqch6:err_sol}, we establish the following closed-form expression for the density of the collective
\begin{equation}\label{eqch6:rho_analytic}
        \rho(x, t) = \bar{\rho}(x)\left[1-\mathrm{exp}(-Kt)\right] + \rho_0(x)\mathrm{exp}(-Kt),
    \end{equation}
    where $\rho_0 = \rho^L_0 + \rho^F_0 + \eta^F_0$.
\end{rem}


\subsection{Design of the reacting term}
Our design of the reacting term $q$ in \eqref{eq:ch6_leaders} and \eqref{eq:ch6_followers} is driven by two main objectives: ($i$) to ensure that the hypothesis about the strict positivity of $\rho^L$ in Theorem \ref{thch6:err_convergence} holds and ($ii$) to achieve a desired leaders-to-followers' mass ratio at steady state  \eqref{eq:mass_ratio_requirement}. 

We achieve these goals choosing the reacting term $q$ as
\begin{equation}\label{eqch6:q}
    q = \frac{1}{2} \left(\rho^L u\right)_x + \frac{1}{2}\left[\rho^* (f*\rho)\right]_x - \frac{D}{2} \rho^*_{xx} + g,
\end{equation}
where 
\begin{align}\label{eqch6:rho_star}
    \rho^* := \rho^L - \rho^F
\end{align}
{is the local imbalance between leaders and followers,} and $g$ obeys the mass action law
\begin{align}\label{eq:g}
    g(x, t) = K_{FL} \rho^F(x, t) - K_{LF} \rho^L(x, t).
\end{align}
The positive reaction rates $K_{FL}$ and $K_{LF}$ represent the propensity of leaders to become followers and of followers to become leaders, respectively.

{
\begin{rem}\label{rem:mass_action_law}
    The reaction term $q$ can be factorized as
    \begin{align}\label{eq:new_react}
        q(x, t) = \kappa_{FL}(x, t) \rho^F(x, t) - \kappa^{LF}(x, t) \rho^L(x, t).
    \end{align}
    This represents a mass action law whose non-constant reaction rates are given by (using \eqref{eqch6:q} and \eqref{eqch6:rho_u})
   \begin{subequations}\label{eq:non_const_reacting_rates}
        \begin{align}
        \kappa_{FL} &= K_{FL} - \frac{Ke}{2\rho^F}+ \frac{D\rho_{xx}}{2\rho^F} - \frac{1}{2\rho^F} \left[\rho(f*\rho)\right]_x,\\
            \kappa_{LF} &= K_{LF}  + \frac{D\rho^*_{xx}}{2\rho^L} - \frac{1}{2\rho^L} \left[\rho^*(f*\rho)\right]_x.
        \end{align}
    \end{subequations}
    Note that this factorization is not unique.
\end{rem}
}



\begin{thm}[Strict positivity of $\rho^L$]\label{thch6:feasibility}
    With $u$ chosen as in \eqref{eqch6:rho_u} and $q$ as in \eqref{eqch6:q}, $\bar{\rho}$ is a steady-state solution of \eqref{eq:behavioral_plasticity_model} with $\bar{\rho}^L, \bar{\rho}^F>0$ and $\bar{\eta}^F\geq 0$ for any $x\in\mathcal{S}$, if and only if
    \begin{equation}\label{eqch6:feasib_cond}
        p > \hat{p} = 1 - \min_x\left[\bar{\rho}(x) \frac{\int_{\mathcal{S}}h(x)\,\mathrm{d}x}{h(x)}\right],
    \end{equation}
    with 
    \begin{align}\label{eqch6:h}
    h(x) = \mathrm{exp}\left[\frac{1}{D}\int (f*\bar{\rho})(x)\,\mathrm{d}x\right].
    \end{align}
\end{thm}

\begin{pf}
    ($\impliedby$) The spatio-temporal dynamics of $\rho^*$ (see \eqref{eqch6:rho_star}) obeys 
 \begin{equation}\label{eqch6:rho_star_dynamics}
    \rho^*_t = 2q - \left(\rho^L u\right)_x - \left[\rho^*(f*\rho)\right]_x + D \rho^*_{xx}.
\end{equation}
    $\rho^L$ and $\rho^F$ can be recovered from $\rho$, $\rho^*$, and $\eta^F$ through the change of variables
 \begin{subequations}\label{eqch6:change_of_variables}
        \begin{align}
        \rho^L &= (1/2)\left[\rho + \rho^* - \eta^F\right],\\
        \rho^F &= (1/2)\left[\rho - \rho^* - \eta^F\right].
    \end{align}
    \end{subequations}
    Substituting \eqref{eqch6:q} into \eqref{eqch6:rho_star_dynamics}, and using \eqref{eqch6:change_of_variables} yields
    \begin{align}\label{eqch6:rho_star_linearized}
        \rho^*_t = -a\,\rho^* + b\,\rho -b\,\eta^F,
    \end{align}
    where {$a$ and $b$ are the constants defined as}
    \begin{align}\label{eq:constants}
            a := K_{FL} + K_{LF},
            \quad b := K_{FL} - K_{LF}.
    \end{align}    
   { Compared to the complex dynamics of $\rho^L$ and $\rho^F$, the dynamics of $\rho$ and $\rho^*$ is linear. Moreover, }under the control in Theorem \ref{thch6:err_convergence}, $\bar{\rho}$ is a steady-state solution for \eqref{eq:ch6_overall_density}, so that we look for steady-state solutions of \eqref{eqch6:rho_star_linearized} and \eqref{eq:ch6_non_react_followers}. We start by considering \eqref{eq:ch6_non_react_followers} with $\eta^F_t = 0$, $\eta^F(x, t) = \bar{\eta}^F(x)$, and $\rho(x, t) = \bar{\rho}(x)$, which gives 
    \begin{align}\label{eqch6:etaF_ss}
        D\bar{\eta}^F_{xx} - \left[\bar{\eta}^F(f*\bar{\rho})\right]_x = 0.
    \end{align}
    Integrating \eqref{eqch6:etaF_ss} twice in space (see Appendix \ref{qppE:ss_solution}) yields
    \begin{align}\label{eqch6:eta_bar_F}
        \bar{\eta}^F(x) = \frac{\Phi^F}{\int_\mathcal{S}h(x)\,\mathrm{d}x} h(x).
    \end{align}
    
    We remark that $\bar{\eta}^F$ is positive, periodic, and $\int_\mathcal{S}\bar{\eta}^F\,\mathrm{d}x = \Phi^F$ by construction (see Appendix \ref{qppE:ss_solution} for more details). We can now find the steady-state of $\rho^*$ by setting $\rho^*_t = 0$, $\rho(x,t)
     = \bar{\rho}(x)$, $\rho^*(x, t) = \bar{\rho}^*(x)$, $\eta^F(x,t) = \bar{\eta}^F(x)$ in \eqref{eqch6:rho_star_linearized}. This gives
    \begin{align}\label{eqch6:rho_star_bar}
        \bar{\rho}^* = (b/a)\,\left[\bar{\rho}-\bar{\eta}^F\right].
    \end{align}
    Hence, using \eqref{eqch6:change_of_variables}, at steady-state we obtain
    \begin{subequations}\label{eq:leaders_followers_ss}
        \begin{align}
            \bar{\rho}^L &= \frac{1}{2} \left[\bar{\rho}\left(1+b/a\right) - \bar{\eta}^F\left(1+b/a\right)\right],\\
            \bar{\rho}^F &=  \frac{1}{2} \left[\bar{\rho}\left(1-b/a\right) - \bar{\eta}^F\left(1-b/a\right)\right].
        \end{align}
    \end{subequations}
    Since $\vert b/a \vert <1$ by construction, $\bar{\rho}^L$ and $\bar{\rho}^F$ are strictly positive if  
    \begin{align}\label{eqch6:rho>eta}
        \bar{\rho}(x)>\bar{\eta}^F(x), \;\forall\,x\in\mathcal{S},
    \end{align}
    which is satisfied under condition \eqref{eqch6:feasib_cond} (substituting \eqref{eqch6:eta_bar_F} into \eqref{eqch6:rho>eta}, and recalling \eqref{eqch6:plasticity}).

    ($\implies$) The existence of a steady-state solution for \eqref{eq:behavioral_plasticity_model} with  $\rho^L, \rho^F >0$ and $\bar{\eta}^F\geq0$ implies that $\bar{\rho}^L + \bar{\rho}^F>0$ $\forall x\in\mathcal{S}$.
    By adding and subtracting $\bar{\eta}^F$, we obtain
    \begin{align}
        \bar{\rho}^L(x) + \bar{\rho}^F(x) + \bar{\eta}^F(x) > \bar{\eta}^F(x), \;\forall\,x\in\mathcal{S},
    \end{align}
    which is equivalent to
    \begin{align}\label{eqch6:final_eq_implies}
        \bar{\rho}(x) > \bar{\eta}^F(x), \;\forall\,x\in\mathcal{S}.
    \end{align}
    Substituting \eqref{eqch6:eta_bar_F} into \eqref{eqch6:final_eq_implies}, and recalling \eqref{eqch6:plasticity}, completes the proof. \qed
\end{pf}
\begin{rem}\label{rem:th2}
    Theorem \ref{thch6:feasibility} gives conditions about the minimum fraction of agents that can switch role such that $\bar{\rho}$ can be a meaningful steady-state solution for \eqref{eq:behavioral_plasticity_model}, that is, \eqref{eq:mass_sum} and \eqref{eq:prob_stat} hold with $\rho^L>0$, $\rho^F, \eta^F \geq 0$. 
\end{rem}
\begin{cor}\label{rem:mass_ratio}
    The requirement in \eqref{eq:mass_ratio_requirement} can be ensured by { appropriately} choosing $K_{LF}$ and $K_{FL}$ in \eqref{eq:g}.
\end{cor}
\begin{pf}
    From \eqref{eq:leaders_followers_ss}, we compute the steady-state leaders-to-followers mass ratio, that is
    \begin{align}   \label{eq:react_rate_choice}
    \frac{\int_\mathcal{S}\bar{\rho}^L(x)\,\mathrm{d}x}{\int_\mathcal{S}\bar{\rho}^F(x)\,\mathrm{d}x} {=\frac{1+b/a}{1-b/a}=\frac{a+b}{a-b}} =  \frac{K_{FL}}{K_{LF}}.
    \end{align}
    Hence, by appropriately choosing the {reaction} rates $K_{LF}$ and $K_{FL}$, we fulfill \eqref{eq:mass_ratio_requirement}.\qed
\end{pf}
\begin{rem}\label{rem:centralized}
    Our control assumes full knowledge of the swarm dynamics. This accommodates swarm robotics scenarios, in which centralized sensing is in place. Moreover, enabling agents' communication to estimate densities allows for distributed implementations \cite{dilorenzo2025decentralized,brumali2025distributed}.
\end{rem}

\begin{exmp}
We illustrate how to use Theorem \ref{thch6:feasibility}, by considering a population of agents interacting via the periodic Morse interaction kernel (long-range attraction, short-range repulsion) 
\begin{equation}\label{eq:morse_kernel}
 f(x) = (1/L_r)\, f_r(x) -(\alpha/L_a) \,f_a(x)   
\end{equation}
where $L_a$ and $L_r$ are the length scales of the attractive and repulsive part of the interaction kernel, and
\begin{equation}
    f_i(x) = \frac{\mathrm{sgn(x)}}{\mathrm{exp}\left(\frac{2\pi}{L_i}\right)-1}\left[\mathrm{exp}\left(\frac{2\pi-\vert x \vert}{L_i}\right)- \mathrm{exp}\left(\frac{\vert x\vert}{L_i}\right)\right],
\end{equation}
with $i=L, R$ (see \cite{boldini_stygmergy} for more details). We set $L_a = \pi$, $\alpha = 2$,  $L_r=\pi/6$, and $D=0.05$. We select a von Mises desired density, that is,
\begin{equation}\label{eq:desired_monomodal}
    \bar{\rho}(x) = Z\,\mathrm{exp}\left[k\cos (x-\mu)\right],
\end{equation}
where we fix mean $\mu=0$ and concentration coefficient $k=1$. $Z$ is chosen so that $\bar{\rho}$ integrates to 1. Using \eqref{eqch6:feasib_cond}, we establish that the fraction of agents allowed to switch role should be larger than $\hat{p}\approx0.15$.
\end{exmp}
\begin{figure*}[t]
     \centering
     \begin{subfigure}[b]{0.22\textwidth}
         \centering
         \includegraphics[width=\textwidth]{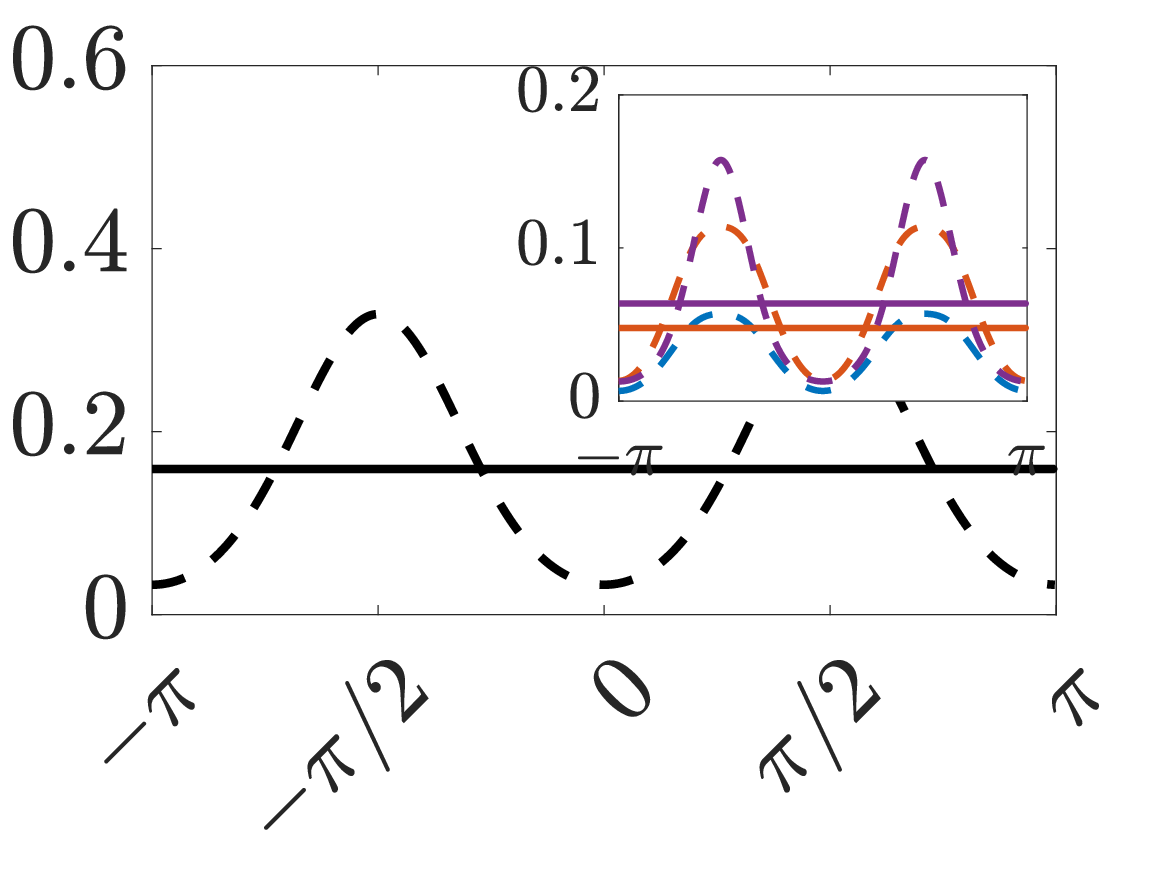}
         \caption{$t=0$}
         \label{sub:react_bimodal_t0}
     \end{subfigure}     
     \begin{subfigure}[b]{0.22\textwidth}
         \centering
         \includegraphics[width=\textwidth]{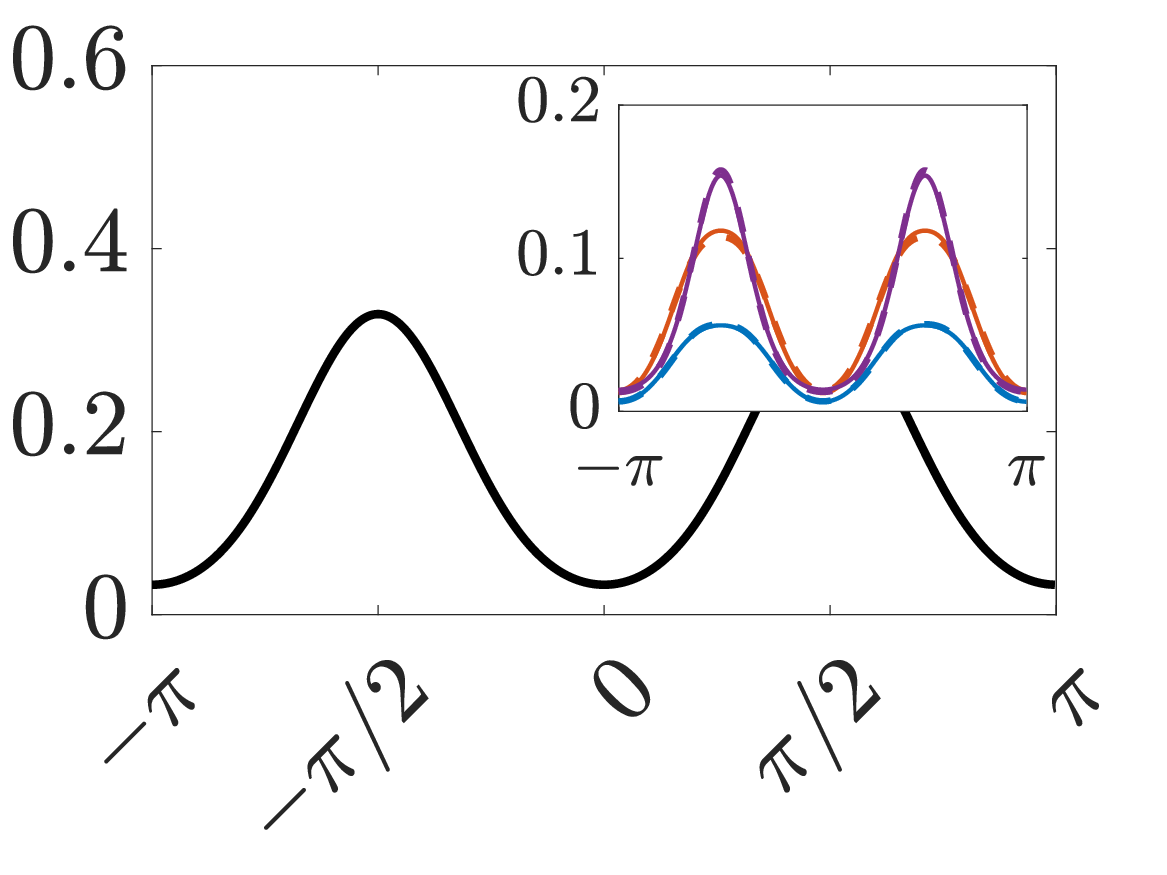}
         \caption{$t=t_\mathrm{f}$}
         \label{sub:react_bimodal_tf}
     \end{subfigure}
     \begin{subfigure}[b]{0.22\textwidth}
         \centering
         \includegraphics[width=\textwidth]{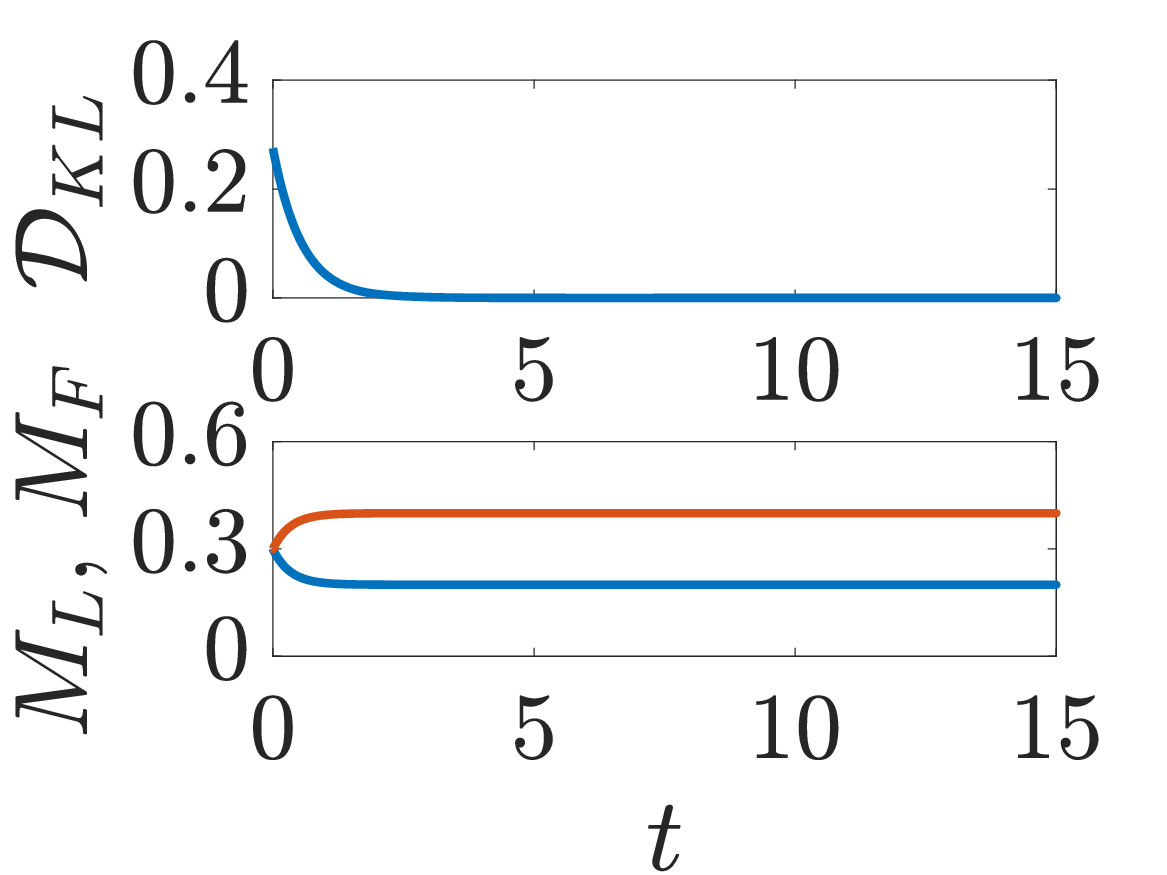}
         \caption{}
         \label{sub:react_bimodal_KL}
     \end{subfigure}     
     \begin{subfigure}[b]{0.22\textwidth}
         \centering
         \includegraphics[width=\textwidth]{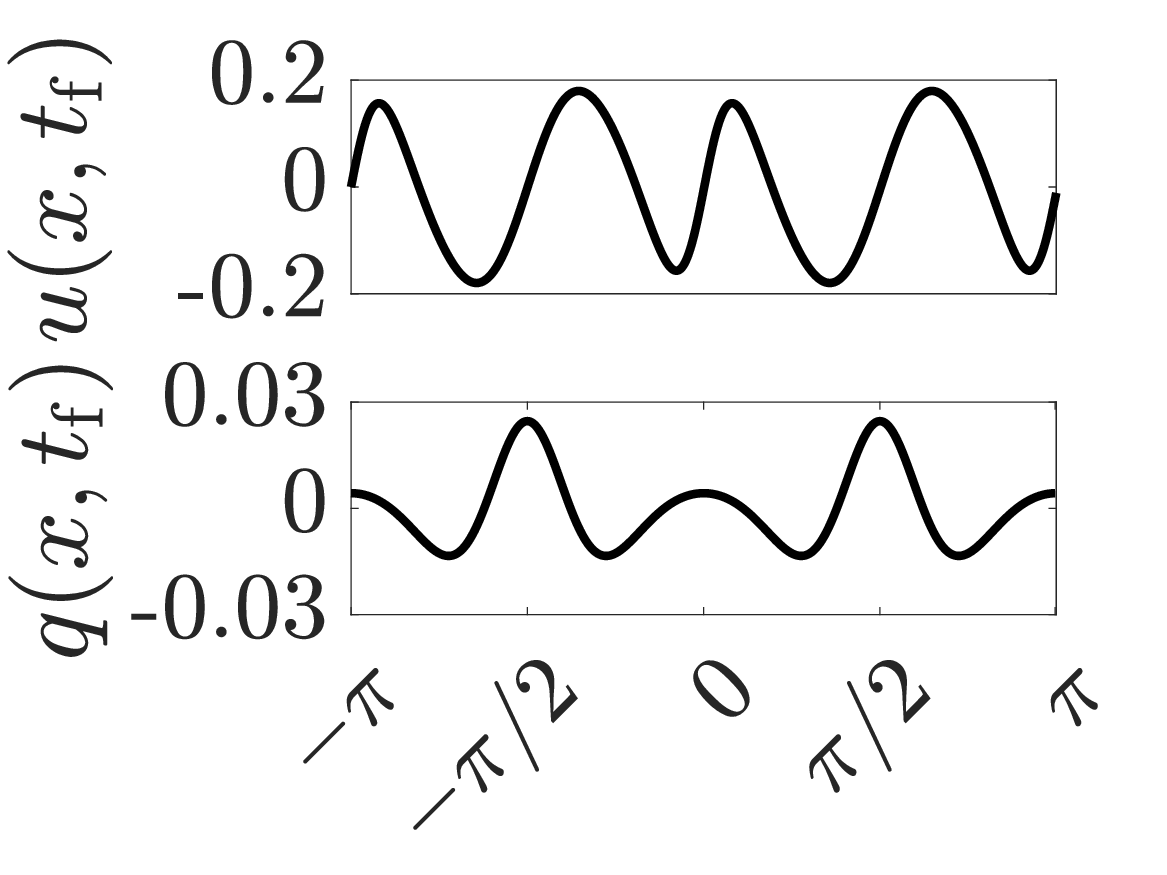}
         \caption{}
         \label{sub:react_bimodal_u_q}
     \end{subfigure}
        \caption{Bimodal regulation. (a,b) Initial/final (solid black) and desired  (dashed black) density of the collective. In the inset, we report the initial/final densities of leaders (solid blue), plastic followers (solid orange), and non-plastic followers (solid purple) along with the density predictions at steady state from Theorem \ref{thch6:feasibility} for the three populations (dashed and same color coding). (c) Time evolution of the KL divergence (top panel) and leaders' and followers' mass (bottom panel). (d) Final distribution profile of the leaders' velocity $u$ (top panel) and reacting term $q$ (bottom panel).}
        \label{fig:reacting_bimodal}
\end{figure*}
\subsection{Stability analysis}
In this section, we assess the stability properties of our control solution, exerted through the velocity field $u$ in \eqref{eqch6:rho_u} and the reactive term $q$ in \eqref{eqch6:q}. From Theorem \ref{thch6:err_convergence}, we know that, if $\rho^L>0$, global convergence of $\rho$ toward $\bar{\rho}$ is ensured. Theorem \ref{thch6:feasibility} instead gives conditions for $\rho^L$ to be striclty positive at steady-state. Hence, we now prove {global} stability of the solution whose existence is proved in Theorem \ref{thch6:feasibility}. Let us recall the function $e = \bar{\rho} - \rho$, and  define error functions
\begin{subequations}
    \begin{align}
        e^*(x, t) &:= \bar{\rho}^*(x) - \rho^*(x, t)\label{eqch6:e_star},\\
        e^\eta(x, t) &:=  \bar{\eta}^F(x) - \eta^F(x, t)\label{eqch6:e_eta},
    \end{align}    
\end{subequations}
with $\bar{\rho}^*$ defined in \eqref{eqch6:rho_star_bar} and $\bar{\eta}^F$ in \eqref{eqch6:eta_bar_F}.

\begin{thm} [{Global} stability]\label{thch6:local_stab}
    Under the conditions of Theorem \ref{thch6:feasibility}, error functions \eqref{eqch6:e}, \eqref{eqch6:e_star}, and \eqref{eqch6:e_eta} { globally} converge to 0 a.e. if 
    \begin{equation}\label{eq:stability_condition}
        \Vert \bar{\rho}_x(\cdot)\Vert_2 <{2D}/{\Vert f(\cdot)\Vert_2}.
    \end{equation}
\end{thm}

\begin{pf}
    The error dynamics under the effect of $u$ and $q$ are given by
   \begin{subequations}
        \begin{align}
            &e_t = -K\, e,\label{eqch6:e_dyn}\\
            &e^*_t = - a \,e^* + b \,e -b\,e^\eta,\label{eqch6:estar_dyn1}\\
            &e^\eta_t + [e^\eta(f*\bar{\rho})]_x - [e^\eta(f*e)]_x = De^\eta_{xx} - [\bar{\eta}^F(f*e)]_x. \label{eqch6:e_eta_dyn1}
        \end{align}
    \end{subequations}
    We substitute \eqref{eqch6:err_sol} into {\eqref{eqch6:e_eta_dyn1}}, yielding
   \begin{multline}\label{eqch6:e_eta_lin}
        e^\eta_t + [e^\eta(f*\bar{\rho})]_x { -\exp(-Kt)\left[e^\eta v^0\right]_x} \\= De^\eta_{xx} {-} \exp(-Kt) \left[\bar{\eta}^F v^0\right]_x,
    \end{multline}
    where $v^0 = f*e^0$ with $e^0(x) = e(x, 0)$.
    
    We introduce the {$\mathcal{L}^2$-norm of the error} $V = \Vert e^\eta\Vert_2^2$, such that its time derivative can be expressed as 
   \begin{multline}\label{eqch6:Vt}
        V_t = 2\int_\mathcal{S} e^\eta e^\eta_t\,dx = 2D\int_\mathcal{S} e^\eta e^\eta_{xx}\,dx \\{-2\exp(-Kt)\int_\mathcal{S}e^\eta \left[\tilde{v}-\left(e^\eta v^0\right)_x\right]dx}\\- 2\int_\mathcal{S} e^\eta\left[e^\eta(f*\bar{\rho})\right]_x\,dx,
    \end{multline}
    where we {used} \eqref{eqch6:e_eta_lin} and $\tilde{v} = (\bar{\eta}^F v^0)_x$
    We expand the first term {on} the right-hand side of \eqref{eqch6:Vt} as
    \begin{align}\label{eqch6:first_term}
        2D\int_\mathcal{S} e^\eta e^\eta_{xx}\,\mathrm{d}x = - 2D\int_\mathcal{S} \left(e^\eta_x\right)^2\,\mathrm{d}x = -2D\Vert e_x^\eta\Vert_2^2,
    \end{align}
    where we applied integration by parts (recalling the periodicity of the functions).
    We similarly expand the {last} term on the right-hand side of \eqref{eqch6:Vt} as
    \begin{multline}\label{eqch6:second_term}
        - 2\int_\mathcal{S} e^\eta\left[e^\eta(f*\bar{\rho})\right]_x\,\mathrm{d}x = 2\int_\mathcal{S} e^\eta_x e^\eta(f*\bar{\rho})(x)\,\mathrm{d}x\\
        = \int_\mathcal{S} \left[\left(e^\eta\right)^2\right]_x(f*\bar{\rho})\,\mathrm{d}x = - \int_\mathcal{S} \left(e^\eta\right)^2(f*\bar{\rho})_x\,\mathrm{d}x,
    \end{multline}
    where we used integration by parts (twice), and exploited the identity $\left(\left(e^\eta\right)^2\right)_x = 2e^\eta e^\eta_x$. { We can apply the same procedure in \eqref{eqch6:second_term} to establish
    \begin{multline}\label{eq:bound_on_exp_term}
        2\exp(-Kt)\int_\mathcal{S} e^\eta \left(e^\eta v^0\right)_x\,\mathrm{d}x \\= \exp(-Kt)\int_\mathcal{S} \left(e^\eta\right)^2v^0_x\,\mathrm{d}x.
    \end{multline}
    }
    Substituting \eqref{eqch6:first_term}, \eqref{eqch6:second_term} and {\eqref{eq:bound_on_exp_term}} into \eqref{eqch6:Vt}, we obtain
    \begin{multline}
        V_t = -2D\Vert e_x^\eta\Vert_2^2 - \int_\mathcal{S} \left(e^\eta\right)^2(f*\bar{\rho})_x\,\mathrm{d}x\\-2\mathrm{exp}(-Kt)\int_\mathcal{S}e^\eta \tilde{v}\,\mathrm{d}x
        { +\mathrm{exp}(-Kt)\int_\mathcal{S}(e^\eta)^2 {v}^0_x\mathrm{d}x}.
    \end{multline}
    By using Poincaré-Wirtinger inequality (see Lemma 2 in \cite{maffettone2024leader}), we can bound {$V_t$} as
    \begin{multline}\label{eqch6:first_bound}
        V_t \leq -2D\Vert e^\eta\Vert_2^2 - \int_\mathcal{S} \left(e^\eta\right)^2(f*\bar{\rho})_x\,\mathrm{d}x\\-2\mathrm{exp}(-Kt)\int_\mathcal{S}e^\eta\tilde{v}\,\mathrm{d}x{ +\mathrm{exp}(-Kt)\int_\mathcal{S}(e^\eta)^2 {v}^0_x\mathrm{d}x}.
    \end{multline}
    For the second term on the right-hand side of \eqref{eqch6:first_bound}, 
    we have
    \begin{multline}\label{eqch6:second_bound}
        \left\vert \int_\mathcal{S} \left(e^\eta\right)^2(f*\bar{\rho})_x\,\mathrm{d}x\right\vert \leq \int_\mathcal{S} \left\vert \left(e^\eta\right)^2(f*\bar{\rho})_x\right\vert\,\mathrm{d}x\\=
        \Vert e^\eta e^\eta(f*\bar{\rho})_x \Vert_1 \leq \Vert e^\eta\Vert_2\Vert e^\eta\Vert_2\Vert (f*\bar{\rho})_x\Vert_\infty\\\leq \Vert e^\eta\Vert_2^2 \Vert f\Vert_2 \Vert \bar{\rho}_x\Vert_2,
    \end{multline}
    where we used  H$\ddot{\mathrm{o}}$lders' inequality, the definition of the derivative of a convolution, and Young's inequality. Similarly, we establish
    {
    \begin{multline}\label{eq:bound_nuovo}
        \left\vert \int_\mathcal{S} \left(e^\eta\right)^2v^0_x\,\mathrm{d}x\right\vert \leq 
        \Vert e^\eta e^\eta v^0_x\Vert_1 \leq \Vert e^\eta\Vert_2^2\Vert v^0_x\Vert_\infty
    \end{multline}
    }
    \vspace{-15pt}
    \begin{multline}\label{eq:third_bound}
        \left\vert-2\mathrm{exp}(-Kt)\int_\mathcal{S}e^\eta \tilde{v}\,\mathrm{d}x\right\vert   \leq 2\mathrm{exp}(-Kt)\Vert e^\eta\tilde{v} \Vert_1 \\\leq 2\mathrm{exp}(-Kt)\Vert e^\eta \Vert_2\Vert\tilde{v} \Vert_2.
    \end{multline}
    Applying bounds \eqref{eqch6:second_bound}, {\eqref{eq:bound_nuovo}} and \eqref{eq:third_bound} to \eqref{eqch6:first_bound}, { yields}
    \begin{multline}\label{eq:final_bound}
        V_t\leq\left(-2D+\Vert f\Vert_2 \Vert \bar{\rho}_x\Vert_2\right)V {+\exp(-Kt)\Vert v^0_x\Vert_\infty V}\\+ 2\mathrm{exp}(-Kt)\Vert \tilde{v} \Vert_{2} \sqrt{V}.
    \end{multline}
    If $\Vert \bar{\rho} \Vert_2 < 2D/\Vert f\Vert_2$, the right-hand side in \eqref{eq:final_bound} converges to 0 thanks to Lemma 4 in \cite{maffettone2024leader} (with {$\beta = 2D-\Vert f\Vert_2 \Vert \bar{\rho}_x\Vert_2$} $\gamma = {\Vert v^0_x\Vert_\infty}$, and $\delta = 2\Vert \tilde{v} \Vert_2$). Hence, by {the} Comparison Lemma \cite{khalil2002nonlinear}, $\Vert e^\eta\Vert_2^2$ converges to 0.

    Since \eqref{eqch6:e_dyn} converges point-wise and \eqref{eqch6:e_eta_dyn1} converges { globally} in $\mathcal{L}^2(\mathcal{S})$,
    we can analyze \eqref{eqch6:estar_dyn1}, rewriting it as
    \begin{align}\label{eq:e*2}
        e^*_t(x, t) = -a e^*(x, t) + w(x, t),
    \end{align}
    where $w$ is a bounded function converging to 0 asymptotically in time and a.e. in $\mathcal{S}$. Computing the unilateral Laplace transform in time to \eqref{eq:e*2} yields
    \begin{align}
        E^*(x, s) = W(x, s)/(s+a).
    \end{align}
    where $E^*$ and $W$ are  Laplace transform of $e^*$ and $w$, respectively. As $a>0$, the final value theorem yields
    \begin{align}
        \lim_{t\to\infty} e^*(x, t) = \lim_{s\to0} \frac{sW(x, s)}{s+a} = 0,
    \end{align}
    where we used the fact that $\lim_{s\to0} sW(x, s) = 0$ since $w$ asymptotically converges to zero. \qed
\end{pf}
%

\section{Numerical Validation}\label{sec:num_valid_1d_React}
{
We validate our claims through two scenarios scoring performance via Kullback-Leibler (KL) divergence in time $\mathcal{D}_{KL} = \int_S \rho\log(\rho/\bar{\rho})\,\mathrm{d}x$. As a measure of steady-state performance, we use the KL divergence at the final instant of our simulations, that is $\mathcal{D}_{KL}^{\mathrm{ss}} = \mathcal{D}_{KL}(t_\mathrm{f})$. For the numerical integration of \eqref{eq:behavioral_plasticity_model}, we use central finite difference in space and forward Euler in time over a mesh of 600 grid points and a time step $\Delta t = 10^{-3}$. 

First, we consider a bimodal von-Mises distribution (two terms as in (37) with $\mu_1=\pi/2$, $\mu_2=-\pi/2$, $k=3$) with interactions driven by a Morse kernel \eqref{eq:morse_kernel} ($L_a=\pi$, $L_r=\pi/2$, $\alpha=2$), $D=0.05$, $K=1$, $K_{FL}=1$, $K_{LF}=2$, $\Phi^F=0.4$, $M^L(0)=M^F(0)=0.3$.} Several messages can be gathered from the results in Fig. \ref{fig:reacting_bimodal}: ($i$) the proposed bio-inspired control scheme is successful in achieving a bimodal density distribution for the collective starting from a uniform one (see Figs.~\ref{sub:react_bimodal_t0} and \ref{sub:react_bimodal_tf}), in agreement with Theorem \ref{thch6:err_convergence} and \ref{thch6:local_stab}; ($ii$) our choice of $\hat{p}$ based on Theorem \ref{thch6:feasibility}, ensures the strict positivity of the steady-state density displacement of the leaders
(see the inset in Fig.~ \ref{sub:react_bimodal_tf}); and ($iii$) our choice of $K_{LF}$ and $K_{FL}$ ensures a steady-state leaders-to-followers mass ratio  $\hat{r} = 1/2$ in agreement with Corollary \ref{rem:mass_ratio}. { We obtained similar results when deliberately violating the sufficient stability condition \eqref{eq:stability_condition} pointing at the conservative nature of our result. These findings are omitted for brevity.}

\begin{figure}[t]
     \centering
     \begin{subfigure}[b]{0.22\textwidth}
         \centering
         \includegraphics[width=\textwidth]{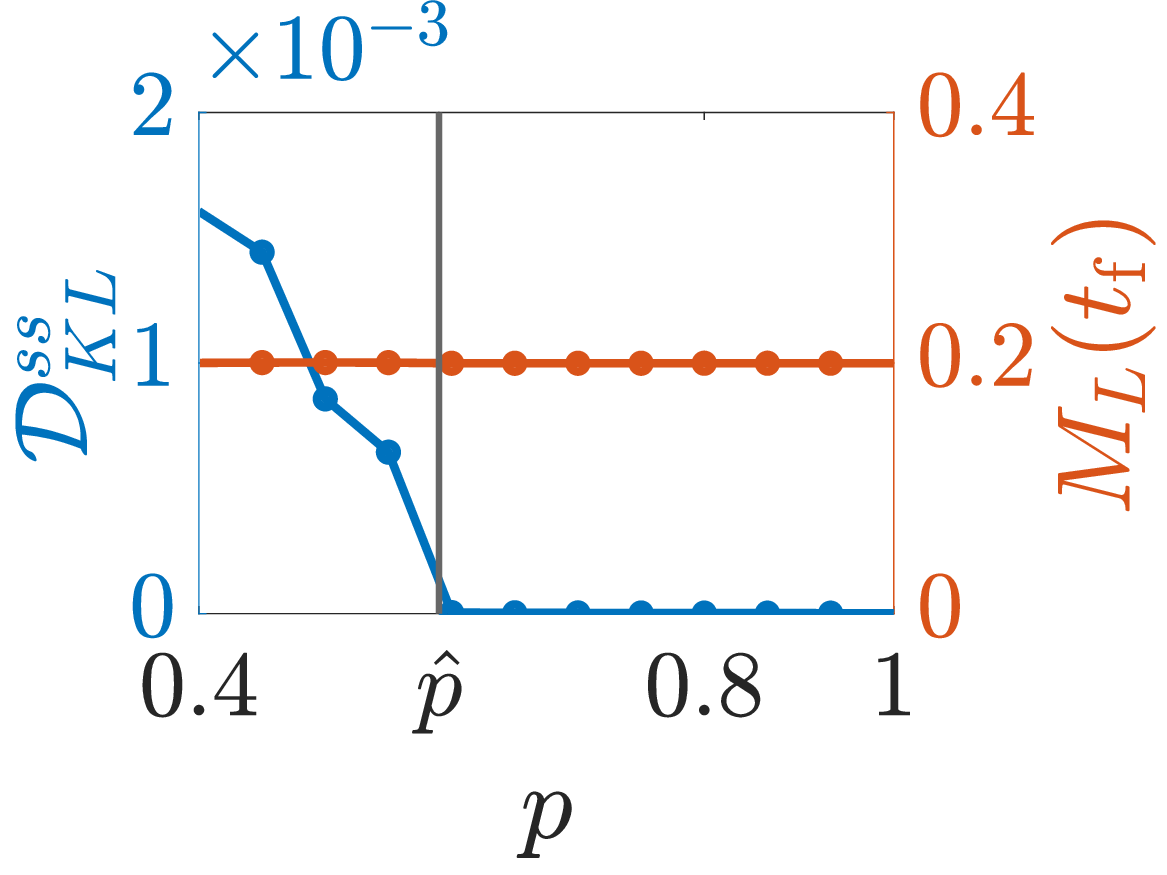}
         \caption{}
         \label{sub:react_rob_diff_coeff}
     \end{subfigure}
     \begin{subfigure}[b]{0.22\textwidth}
         \centering
         \includegraphics[width=\textwidth]{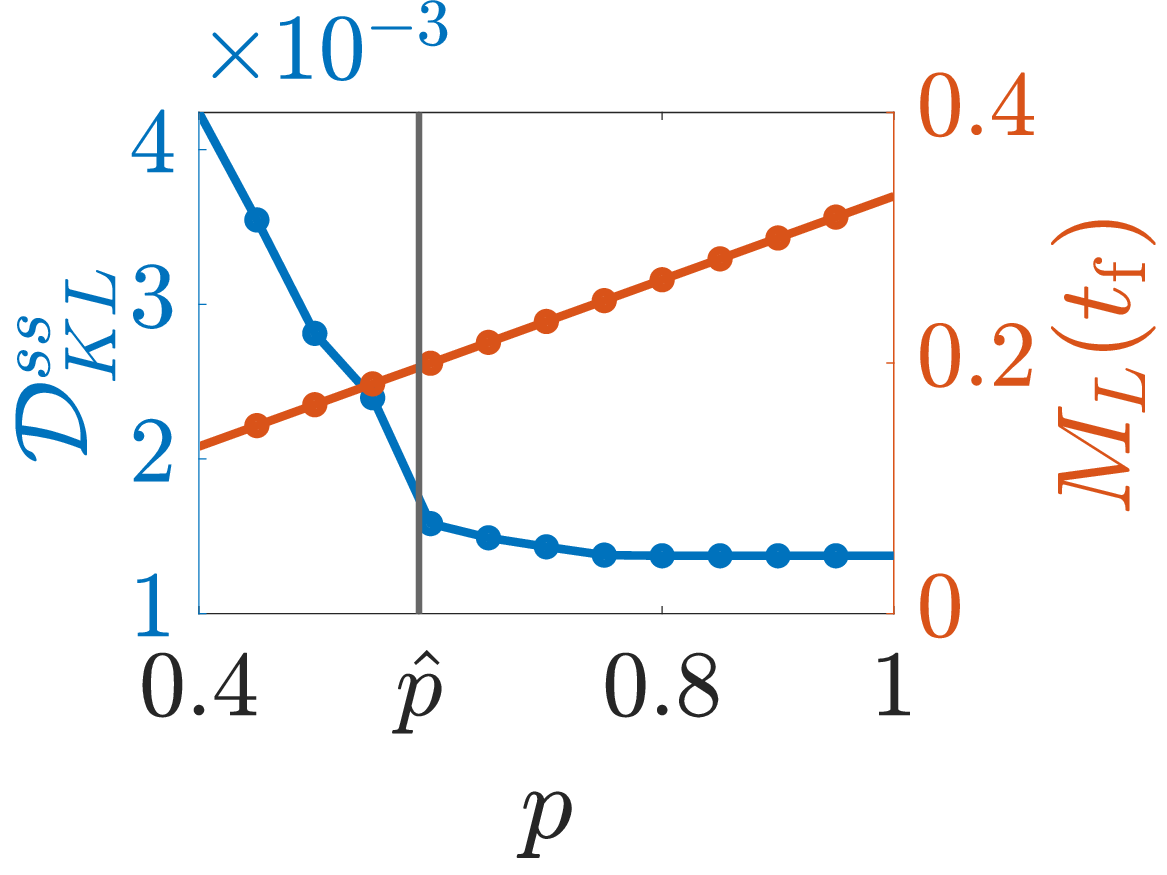}
         \caption{}
         \label{sub:kernel_uncertainties}
     \end{subfigure}
        \caption{Robustness analysis to perturbations in (a) diffusion coefficients and (b) parameters of the interaction kernels. For different values of $p$, we show $\mathcal{D}_{KL}^{ss}$ (blue) and leaders' mass (orange) at steady-state (in solid gray the predicted minimum plasticity { from Theorem \ref{thch6:feasibility}}).}
        \label{fig:reacting_robustness}
\end{figure}

Next, we demonstrate how the fraction of plastic agents affects robustness. We consider a desired von Mises distribution ($\kappa=1$, $\mu=0$), Morse interactions as in \eqref{eq:morse_kernel}($L_a = \pi$, $L_r = \pi/4$, $\alpha = 2$), $K=10$, and $D = 0.02$.
We introduce perturbations to either the diffusion coefficient -- doubling it for the followers in Eqs.~\eqref{eq:ch6_non_react_followers} and \eqref{eq:ch6_followers} with respect to its nominal value used for the leaders in \eqref{eq:ch6_leaders}-- or the interaction kernel parameters -- reducing $L_a$ by 20\% and increasing $L_r$ by 20\% with respect to their nominal values for all the agents. Starting from equilibrium configurations\footnote{For $p<\hat{p}$ (see Theorem \ref{thch6:feasibility}), steady-state configurations $\bar{\rho}^L$ and $\bar{\rho}^F$ are negative in some regions. For these cases, we translate initial configurations upwards to become non-negative and re-normalize them to a predefined mass.}, we assess performance degradation for different values of $p$. When the perturbation affects only followers (as for the test with respect to $D$), we choose $K_{LF}$ and $K_{FL}$ to ensure the steady-state leaders' mass is constant across different values of $p$. { This ensures that possible improved robustness properties are not due to an increase in the amount of leaders at steady-state.}
Results in Figs. \ref{sub:react_rob_diff_coeff} and  \ref{sub:kernel_uncertainties} show that above the minimum  threshold for $p$ predicted in Theorem \ref{thch6:feasibility}, agents rearrange to counteract perturbations, maintaining steady-state performance. Below this threshold, performance degrades significantly.

\section{Agent-based Model}\label{sec:abm}
To bridge the gap between our continuum framework and practical implementation, we develop an agent-based model capturing essential PDE dynamics at the individual level. 
{Following the approach in \cite{albi2024kinetic}, where density dynamics with transient leaders are considered, the model consists of coupled stochastic differential equations with Gaussian noise and stochasticity due to roles' switching}
\begin{subequations}\label{eq:agent-based-equations}
    \begin{multline}
        \mathrm{d}x_i = \Bigg[\frac{1}{N^{LF}} \sum_{j=1}^{N^{LF}} f(\{x_i, x_j\}) + \frac{1}{M}\sum_{j=1}^M f(\{x_i, y_j\})\bigg]\mathrm{d} t\\+ u_i\lambda_i\mathrm{d}t+\sqrt{2D}\,\mathrm{d}W_i{}, \;i=1, \dots, N^{LF},\label{eq:leaders_follwers_foll_abm}
    \end{multline}
    \vspace{-15pt}
    \begin{multline}
        \mathrm{d}y_i = \Bigg[\frac{1}{M} \sum_{j=1}^M f(\{y_i, y_j\}) + \frac{1}{N^{LF}} \sum_{j=1}^{N^{LF}} f(\{y_i, x_j\})\Bigg]\mathrm{d}t \\+\sqrt{2D}\,\mathrm{d}W_i{}, \;i=1, \dots, M.\label{eq:non_plastic_foll_abm}
    \end{multline}
\end{subequations}

Here, $x_i$ are positions of leaders and plastic followers, $y_i$ are positions of non-plastic followers, $N^{LF}$ and $M$ are their respective numbers, and $W_i$ is a standard Wiener process. Control $u_i(t) = u(x_i, t)$ is computed via spatial sampling, and $\lambda_i \in \{0, 1\}$ indicates if agent $i$ is a leader ($\lambda=1$) or follower ($\lambda=0$). $\lambda_i$ is updated stochastically with rates $\kappa_{LF}(x_i, t)$ and $\kappa_{FL}(x_i, t)$ {-- see \eqref{eq:non_const_reacting_rates}}.

\begin{figure}[t]
     \centering
     \begin{subfigure}[b]{0.22\textwidth}
         \centering
         \includegraphics[width=\textwidth]{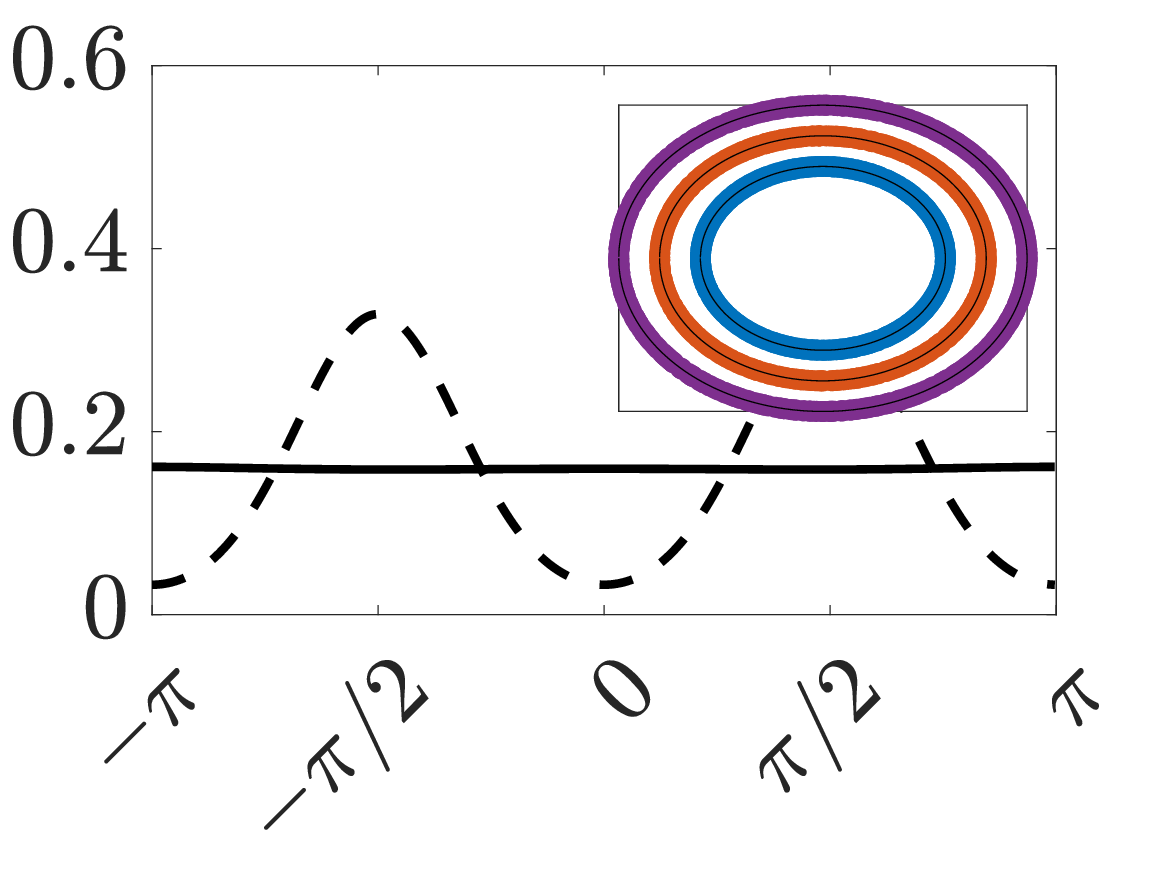}
         \caption{$t=0$}
         \label{sub:react_monomod_t0_disc}
     \end{subfigure}     
     \begin{subfigure}[b]{0.22\textwidth}
         \centering
         \includegraphics[width=\textwidth]{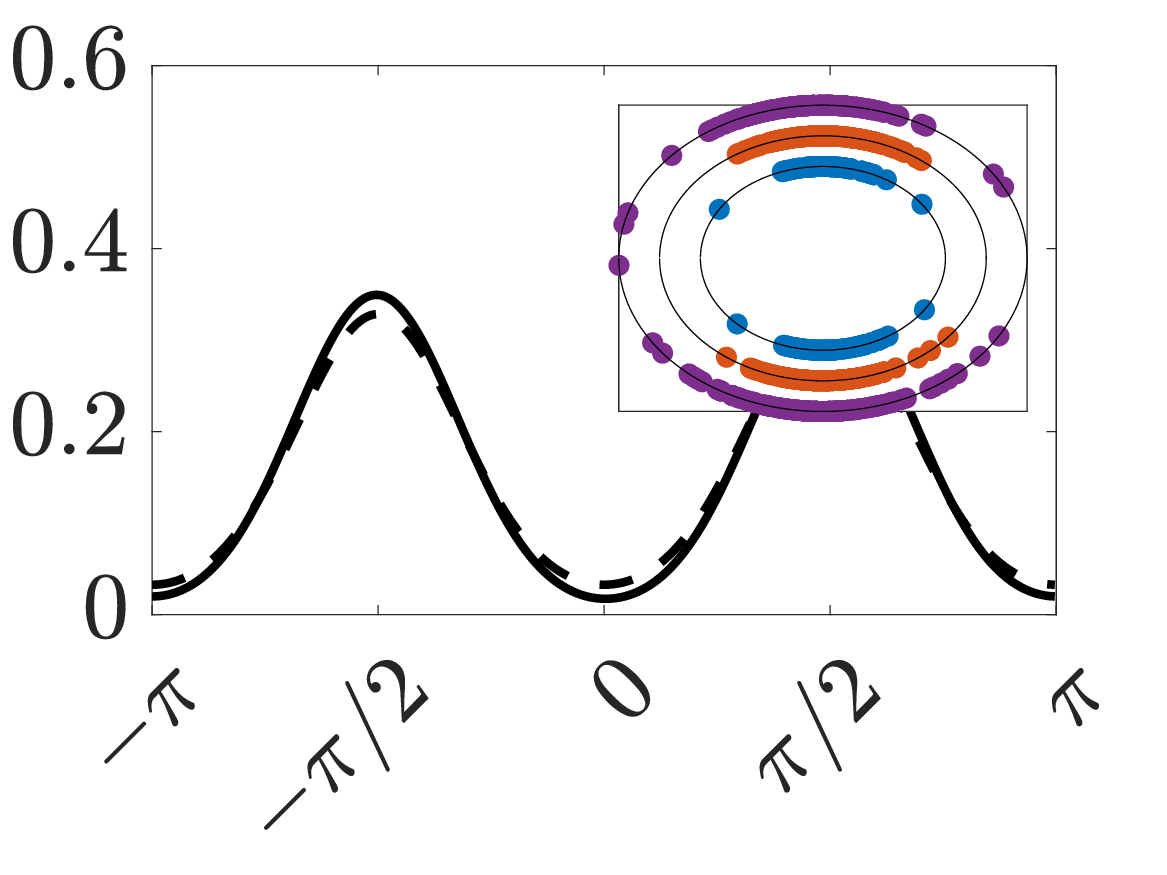}
         \caption{$t=t_\mathrm{f}$}
         \label{sub:react_bimodal_t1_disc}
     \end{subfigure}
     \begin{subfigure}[b]{0.22\textwidth}
         \centering
         \includegraphics[width=\textwidth]{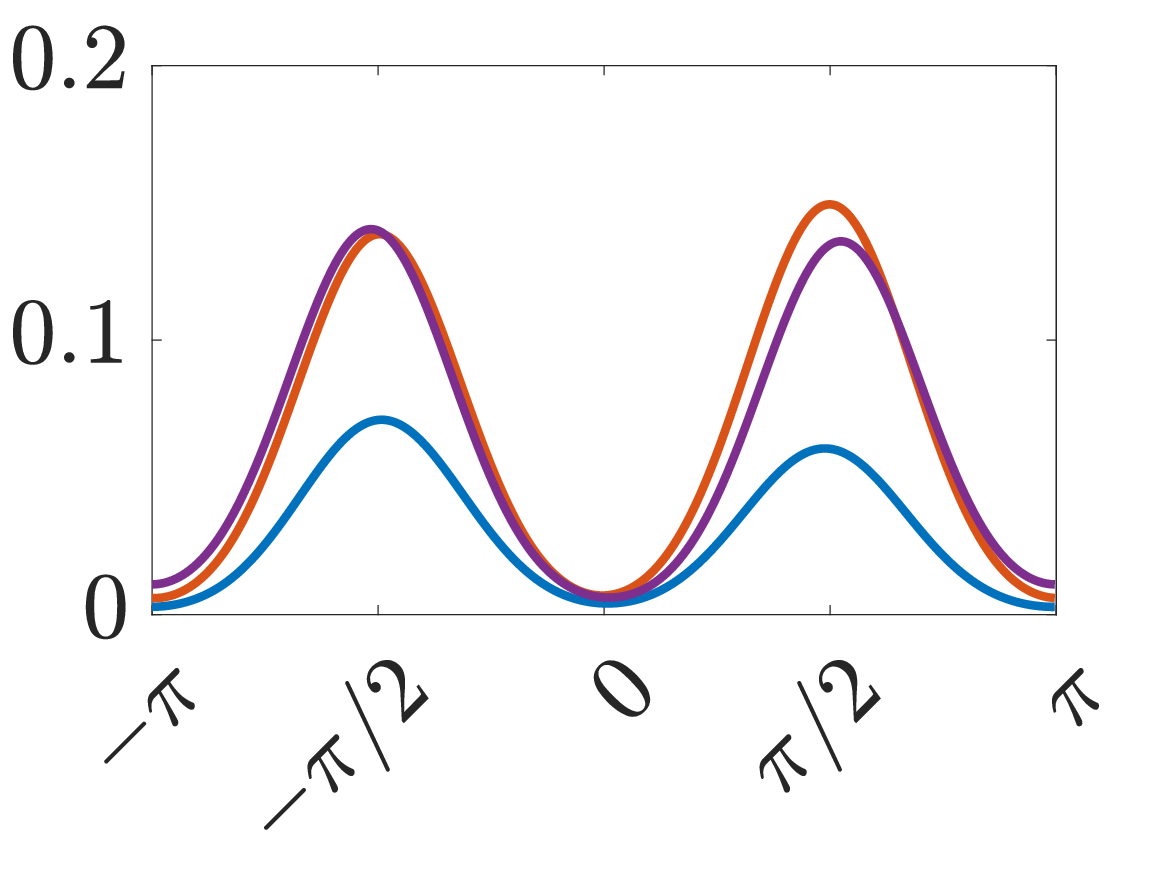}
         \caption{$t=t_\mathrm{f}$}
         \label{sub:react_bimodal_final_densities}
     \end{subfigure} 
     \begin{subfigure}[b]{0.22\textwidth}
         \centering
         \includegraphics[width=\textwidth]{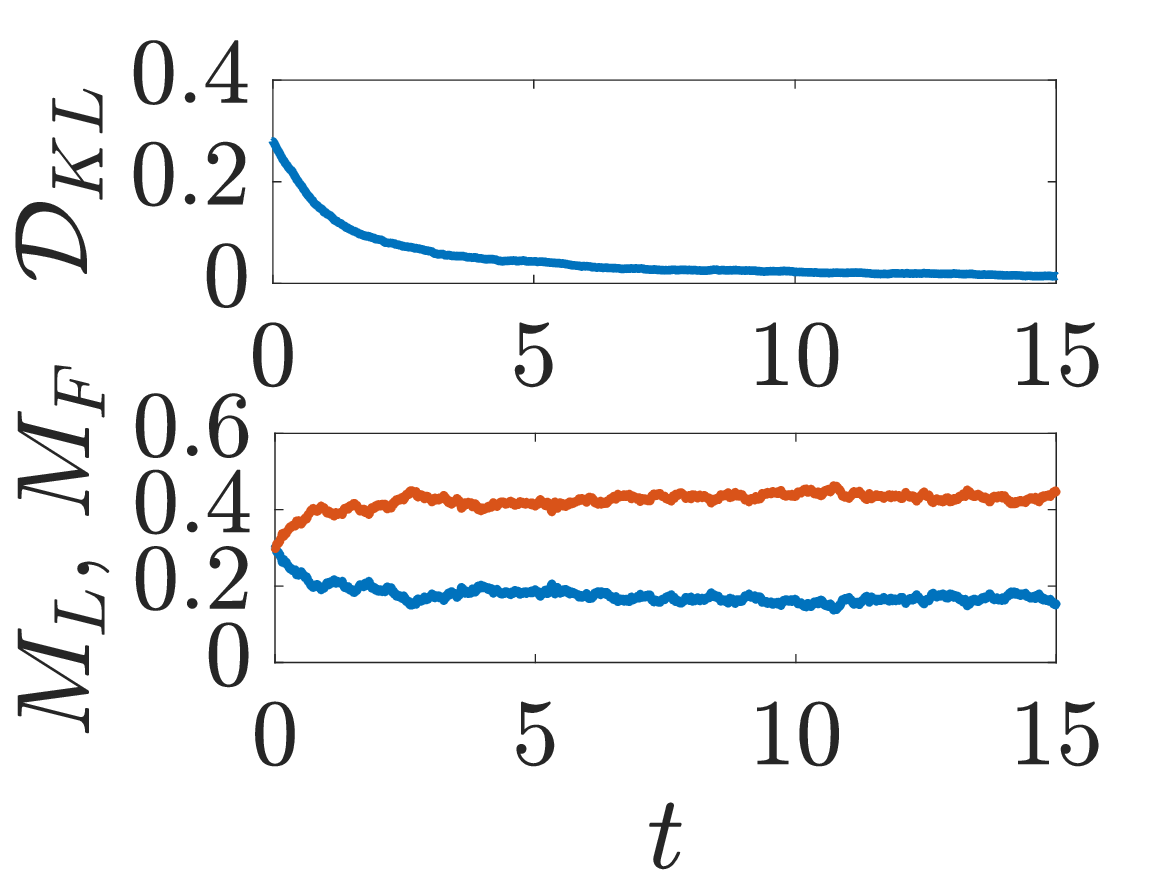}
         \caption{}
         \label{sub:react_bimodal_dkl}
     \end{subfigure}  
        \caption{Agent-based bimodal regulation. (a, b) Initial/final collective densities (solid black). In the inset, discrete displacement of agents: leaders (blue), plastic followers (orange), non-plastic followers (purple), plotted on concentric circles for visualization. (c) Steady-state densities of the three populations. (d) Time evolution of KL divergence (top) and leaders' and followers' mass (bottom).} 
        \label{fig:reacting_bimodal_disc}
\end{figure}

We consider a setup analogous to Sec. \ref{sec:num_valid_1d_React} with $N=1000$ agents (300 initial leaders, 300 plastic followers, 400 non-plastic followers), using Euler-Maruyama integration with $\Delta t = 10^{-3}$, $t_\mathrm{f} = 15$. Control inputs are computed via kernel density estimation using Matlab's \verb|circ_ksdensity| with \verb|msn| method \cite{silverman2018density} (width 50, 600 mesh points). Parameters are set to $D=0.05$, $K=1$, $K_{FL} = 2$, $K_{LF} = 1$.

Fig. \ref{fig:reacting_bimodal_disc} shows a realization of \eqref{eq:agent-based-equations}. Performance degradation compared to the continuum case (Fig. \ref{fig:reacting_bimodal}) arises from finite-size effects. The steady-state leaders-to-followers ratio is approximately 1/2, consistent with Corollary \ref{rem:mass_ratio}. Over 50 stochastic realizations, we obtain average {$\mathcal{D}_{KL}^{\mathrm{ss}}$} $0.02$ ($\pm 0.01$) and average leaders-to-followers mass ratio $0.48$ ($\pm 0.1$).

\section{Higher-Dimensional Extension}\label{sec_ch6:higher_dim}
Next, we extend the theoretical framework to periodic domains in higher-dimensions, $\Omega = [-\pi, \pi]^d$ ($d=2, 3$). The model in \eqref{eq:behavioral_plasticity_model} becomes
\begin{subequations}
    \begin{align}
        \rho^L_t + \nabla \cdot [\rho^L \mathbf{u} + \rho^L(\mathbf{f}*\rho)]  &= D\nabla^2\rho^L + q,\\
        \rho^F_t + \nabla \cdot [\rho^F(\mathbf{f}*\rho)] &= D\nabla^2\rho^F - q,\\
        \eta^F_t + \nabla \cdot [\eta^F(\mathbf{f}*\rho)] &= D\nabla^2\eta^F,\label{eqch6:static_foll_d_dim}
    \end{align}
\end{subequations}
where $\mathbf{f}$ is a $d$-dimensional periodic kernel and $\rho = \rho^L+\rho^F+\eta^F$. The system is complemented with periodic boundary conditions and initial conditions similar to \eqref{eq:behavioral_plasticity_model}. 

\subsection{Bio-inspired Control}
\begin{thm}
    Assume $\rho^L > 0$ for any $x\in\Omega$ and $t\in\mathbb{R}_{0}$. Choosing 
    \begin{multline}\label{eqch6:div_rel}
        \nabla \cdot \left(\rho^L\mathbf{u}\right) = - K e(\mathbf{x}, t) - \nabla\cdot \left[\rho(\mathbf{f}*\rho)\right]{+} D\nabla^2\rho
    \end{multline}
    where $K>0$ is a control gain, the error dynamics converges globally and exponentially to 0 pointwise in $\Omega$.
\end{thm}

\begin{pf}
    The error dynamics obeys the higher-dimensional extension of \eqref{eqch6:err_dyn}. Choosing \eqref{eqch6:div_rel} {transforms} the error dynamics {into} \eqref{eq:err_lin}, proving the claim. \qed
\end{pf}

\begin{rem}\label{rem:u_from_Y}
To uniquely recover $\mathbf{u}$, a vector field, from the scalar relation \eqref{eqch6:div_rel}, extra constraints need to be included, {similar} to \cite{maffettone2023hybrid}. Specifically, we set $\mathbf{w} := \rho^L \mathbf{u}$, and 
\begin{align}
Y = - K e - \nabla\cdot \left[\rho(\mathbf{f}*\rho)(\mathbf{x}, t)\right]{+} D\nabla^2\rho,
\end{align}
so that we can pose the problem 
\begin{equation}
    \begin{cases}
        \nabla \cdot \mathbf{w} = Y,\\
        \nabla \times \mathbf{w} = 0,
\end{cases}
\end{equation}
where we added a zero-curl condition to \eqref{eqch6:div_rel}. Such a problem is analogous to the Poisson equation $\nabla^2 \varphi = -Y$, where $\mathbf{w} = -\nabla\varphi$, { and can be solved } using Fourier series expansion. {Once $\mathbf{\varphi}$ is recovered, $\mathbf{u} = -\nabla\varphi / \rho^L$.}
\end{rem}
Analogously to the one-dimensional case, the resulting control velocity field $\mathbf{u}$ is well defined only if $\rho^L$ is strictly positive. 
Such a constraint can be ensured by appropriately choosing the reacting function $q$. In particular, extending \eqref{eqch6:q} to higher dimensions, we establish
\begin{align}\label{eq:q_hd}
    q =  \frac{1}{2}\, \nabla \cdot \left(\rho^L \mathbf{u}\right) + \frac{1}{2}\,\nabla\cdot \left[\rho^* (\mathbf{f}*\rho)\right] - \frac{D}{2} \nabla^2\rho^* + g,
\end{align}
where $\rho^* = \rho^L - \rho^F$ and $g$ is the mass action law in \eqref{eq:g}.

\subsection{Stability Analysis}
\begin{thm}\label{th:feasibility_hd}
    Assume the interaction kernel to be isotropic, that is,
     \begin{multline}\label{eq:isotropic_kernel}
        \int (f_1*\psi)(x_1, x_2, x_3) \,\mathrm{d}x_1 =  \int (f_2*\psi) (x_1, x_2 x_3)\,\mathrm{d}x_2\\=  \int (f_3*\psi) (x_1, x_2, x_3)\,\mathrm{d}x_3,
    \end{multline}
    for any periodic $\psi$. Choosing $\mathbf{u}$ according to \eqref{eqch6:div_rel} (see Rem. \ref{rem:u_from_Y}) and $q$ as in \eqref{eq:q_hd} implies that $\bar{\rho}$ is a steady-state solution for the dynamics of $\rho$, with $\rho^L, \rho^F >0$, and $\eta^F\geq 0$, if and only if the higher-dimensional extension of \eqref{eqch6:feasib_cond} holds.
\end{thm}

\begin{pf}
Under the additional assumption of isotropic kernel, the proof follows the same steps of those in Theorem \ref{thch6:feasibility}. The only difference is the computation of the steady-state solution of $\eta^F$ (see \eqref{eqch6:eta_bar_F}). Setting $\eta^F_t = 0$ and $\rho = \bar{\rho}$  in \eqref{eqch6:static_foll_d_dim}, we obtain
\begin{align}
    \nabla \cdot \left[\bar{\eta}^F(\mathbf{f}*\bar{\rho})\right] &= D\nabla^2\bar{\eta}^F,
\end{align}
which is rewritten as
\begin{align}\label{eqch6_eta_d_dim_pass}
    \nabla \cdot \left[\bar{\eta}^F(\mathbf{f}*\bar{\rho}) - D\nabla\bar{\eta}^F\right] = 0.
\end{align}
Equation \eqref{eqch6_eta_d_dim_pass} is fulfilled if
%
\begin{align}\label{eqch6_eta_d_dim_pass2}
    \nabla\bar{\eta}^F = ({1}/{D})\,\bar{\eta}^F(\mathbf{f}*\bar{\rho}).
\end{align}
Equation \eqref{eqch6_eta_d_dim_pass2} is a vectorial differential relation involving the partial derivatives of the scalar unknown $\bar{\eta}^F$, thus resulting in the ill-posed problem
\begin{align}\label{eqch6:sys}
\begin{cases}
\bar{\eta}^F_{x_1}(x_1, x_2) = \frac{1}{D} \bar{\eta}^F(x_1, x_2)(f_1*\bar{\rho})(x_1, x_2),\\
\bar{\eta}^F_{x_2}(x_1, x_2) = \frac{1}{D} \bar{\eta}^F(x_1, x_2)(f_2*\bar{\rho})(x_1, x_2).
\end{cases}
\end{align}
Here, without loss of generality, we set $d=2$, $\mathbf{f} = [f_1, f_2]$, $\mathbf{x} = [x_1, x_2]$ (the case $d=3$ is a trivial extension). We can now solve the two components of \eqref{eqch6:sys} separately, and check under which conditions they are equal.   

By solving the first component of \eqref{eqch6:sys}, we establish
\begin{equation}\label{eqch6:eta_f_bar_in}
    \bar{\eta}^F(x_1, x_2) = C_1(x_2) \mathrm{exp} \left\{ \frac{1}{D} \int (f_1*\bar{\rho})(x_1, x_2)\,\mathrm{d}x_1\right\}.
\end{equation}
where $C_1$ is a function of $x_2$ resulting from the spatial integration with respect to $x_1$.
Similarly, if we integrate the second equation of \eqref{eqch6:sys}, we get
\begin{equation}\label{eqch6:eta_f_bar_x2}
    \bar{\eta}^F(x_1, x_2) = C_2(x_1) \mathrm{exp} \left\{ \frac{1}{D} \int (f_2*\bar{\rho})(x_1, x_2)\,\mathrm{d}x_2\right\}.
\end{equation}
where $C_2$ is a function of $x_1$ resulting from the spatial integration with respect to $x_2$.

For \eqref{eqch6:eta_f_bar_in} and \eqref{eqch6:eta_f_bar_x2} to be equal, $C_1(x_2) = C_2(x_1) = C$, and the isotropic hypothesis \eqref{eq:isotropic_kernel} must hold. 
The value of $C$ can finally be chosen so that $\bar{\eta}^F$ integrates to $\Phi^F$ (note that the steady-state solution of \eqref{eqch6:static_foll_d_dim} is in the same form of its 1D counterpart - see \eqref{eqch6:eta_bar_F}). The remainder of the proof follows that of Theorem \ref{thch6:feasibility}.\qed
\end{pf}

\begin{rem}\label{rem:scaling_to_higher_dim}
  Many interaction kernels in the literature satisfy condition \eqref{eq:isotropic_kernel}. Under this hypothesis, the steady-state solution $\bar{\eta}^F$ is uniquely defined, since $\bar{\eta}^F(\mathbf{f}*\bar{\rho})-D\nabla\bar{\eta}^F = -\nabla\Psi$, where $\Psi$ is a harmonic scalar potential. { This is analogous to couple \eqref{eqch6_eta_d_dim_pass} with $\nabla\times\left[\bar{\eta}^F (\mathbf{f}*\bar{\rho})-D\nabla\bar{\eta}^F\right] = 0$.}
\end{rem}

Next, we extend Theorem \ref{thch6:local_stab} to higher dimensions.

\begin{figure*}[t]
     \centering
     \begin{subfigure}[b]{0.22\textwidth}
         \centering
         \includegraphics[width=\textwidth]{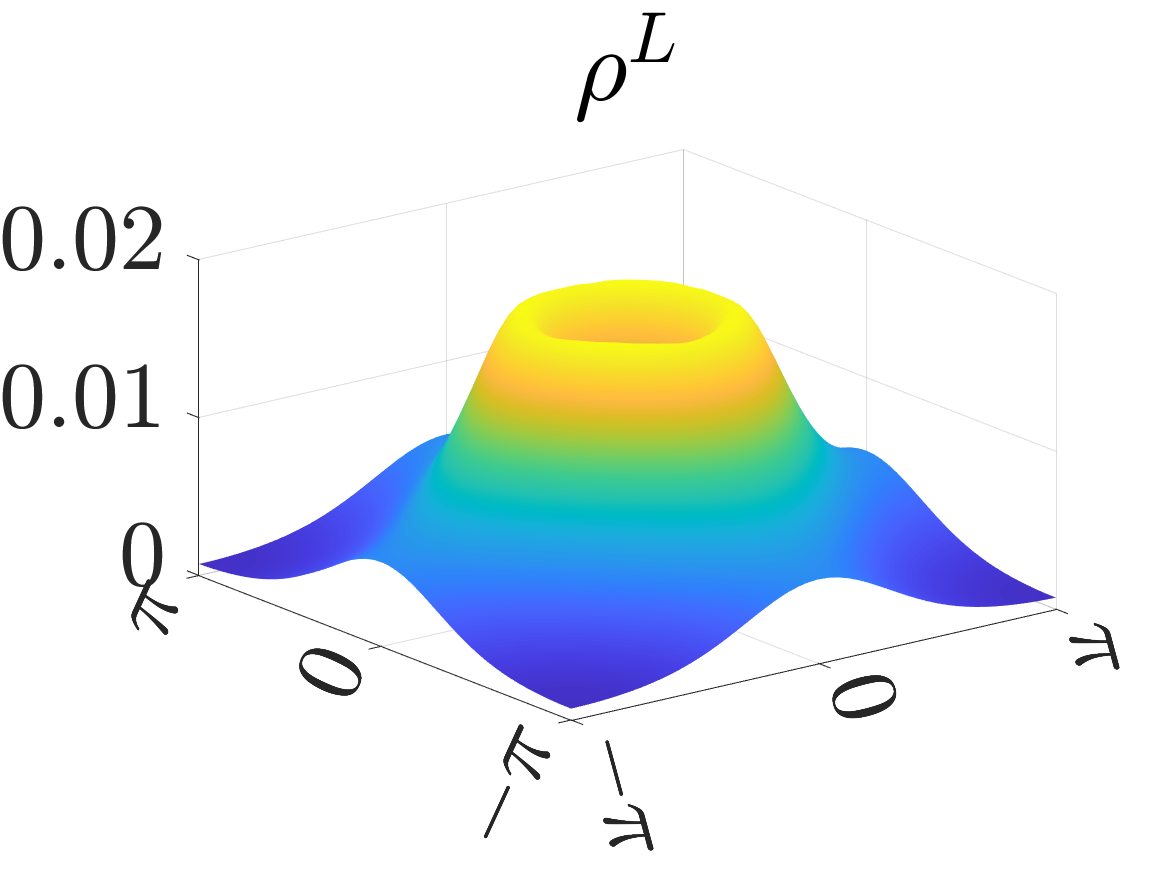}
         \caption{$t=t_\mathrm{f}$}
         \label{sub:react_monomod_2d_rho_l}
     \end{subfigure}     
     \begin{subfigure}[b]{0.22\textwidth}
         \centering
         \includegraphics[width=\textwidth]{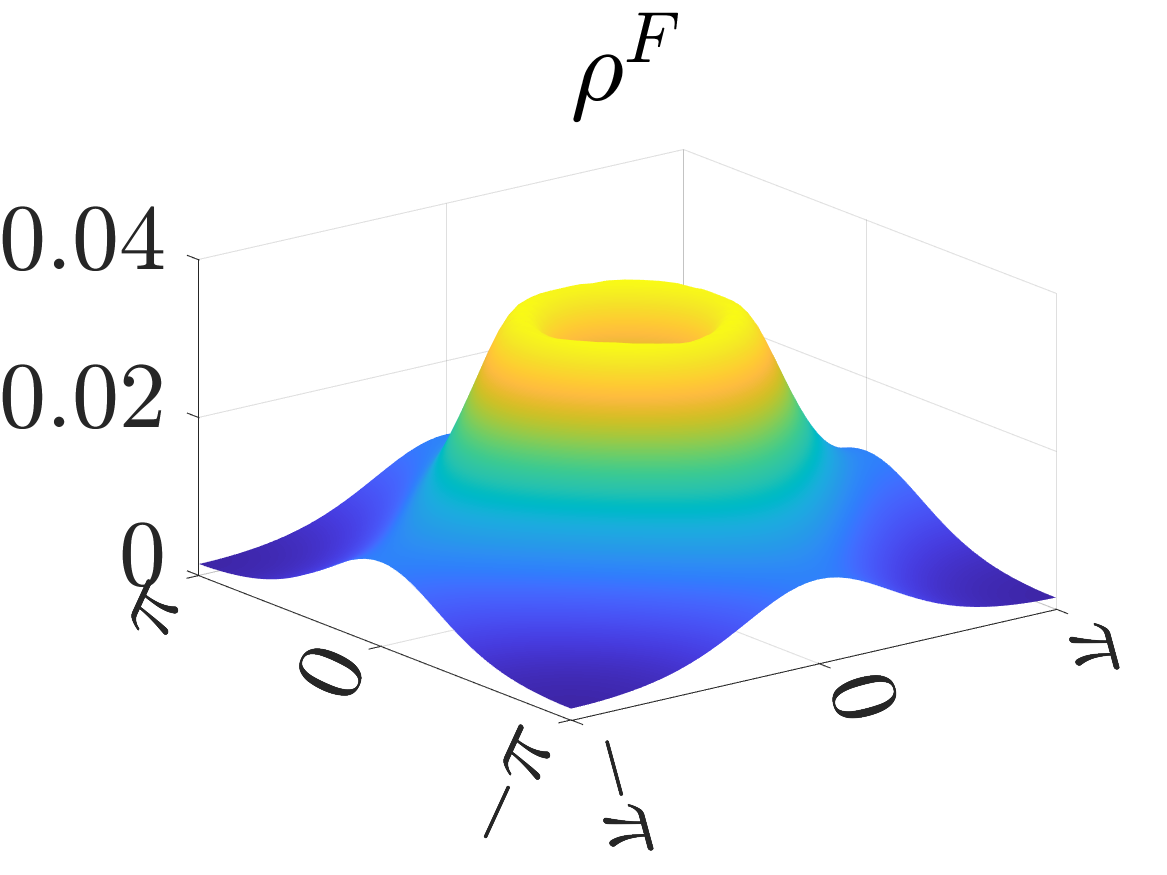}
         \caption{$t=t_\mathrm{f}$}
         \label{sub:react_monomod_2d_rho_f}
     \end{subfigure}
     \begin{subfigure}[b]{0.22\textwidth}
         \centering
         \includegraphics[width=\textwidth]{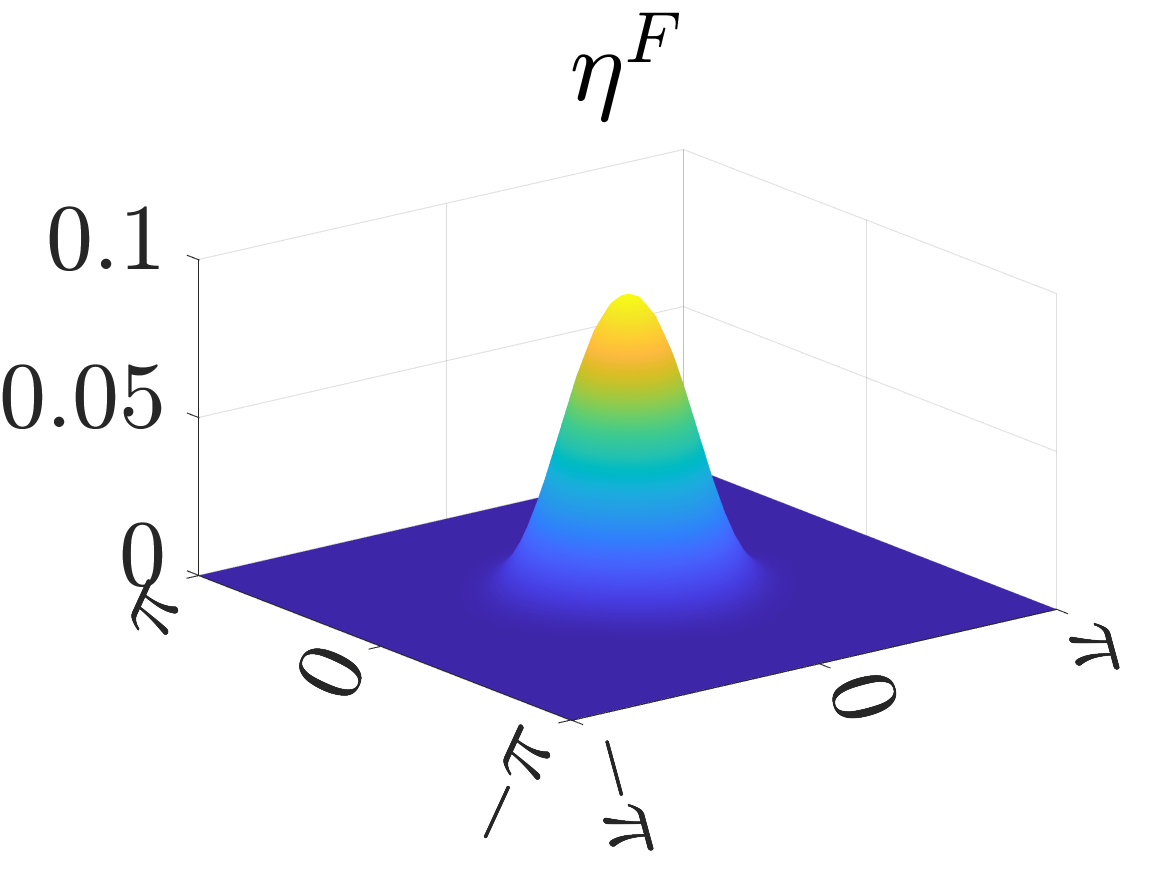}
         \caption{$t=t_\mathrm{f}$}
         \label{sub:react_monomod_2d_eta_f}
     \end{subfigure}     
     \begin{subfigure}[b]{0.22\textwidth}
         \centering
         \includegraphics[width=\textwidth]{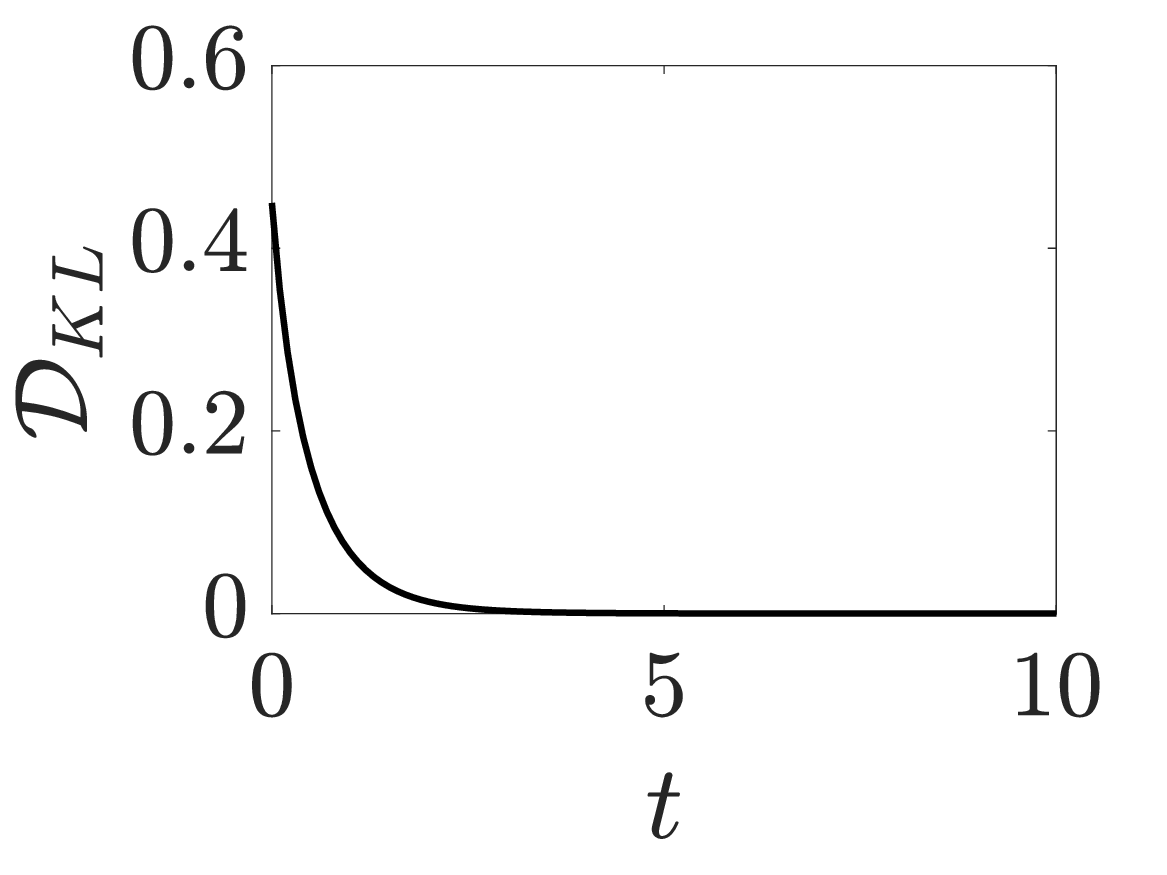}
         \caption{}
         \label{sub:react_monomod_kl_div_2D}
     \end{subfigure}
        \caption{2D Mono-modal regulation. Final densities of (a) leaders', (b) plastic followers', and (c) non-plastic followers. (d) Time evolution of the KL divergence between $\rho$ and $\bar{\rho}$ (truncated to 10 time units for visualization purposes).}
        \label{fig:reacting_monomodal_2d}
\end{figure*}

\begin{thm}
    Under the conditions of Theorem \ref{th:feasibility_hd} and if 
    \begin{equation}
        \sum_{i=1}^d \Vert\bar{\rho}_{x_i}(\cdot) \Vert_2 \Vert f_i(\cdot)\Vert_2 < 2D,
    \end{equation}
    the error dynamics { globally} converges to 0 in $\mathcal{L}^2(\Omega)$.
\end{thm}

\begin{pf}
The proof follows the same structure of that of Theorem \ref{thch6:local_stab}. The time derivative of { $V=\Vert e^\eta\Vert_2^2$} can be rewritten as 
\begin{multline}\label{eq:Vt_hd}
V_t = -2D\Vert\nabla e^\eta \Vert_2^2 - \int_\Omega \left(e^\eta\right)^2\nabla\cdot\left(\mathbf{f}*\bar{\rho}\right)\,\mathrm{d}\mathbf{x} \\-2\mathrm{exp}(-Kt) \int_\Omega e^\eta \tilde{\mathbf{v}} \,\mathrm{d}\mathbf{x} {+\exp(-Kt) \int_{\Omega} (e^\eta)^2 \nabla\cdot \mathbf{v^0}\,\mathrm{d}\mathbf{x}},
\end{multline}
using the divergence theorem, vectorial identities, and posing {$\mathbf{v}^0 = \mathbf{f}*e^0$ and $\tilde{\mathbf{v}} = \nabla \cdot(\bar{\eta}^F\,\mathbf{v}^0 )$}. The first term on the right-hand side can be bounded using {the} Poincaré-Wirtinger inequality, and for the second one the following bound holds:
\begin{multline}
    \left\vert \int_\Omega \left(e^\eta\right)^2\nabla\cdot\left(\mathbf{f}*\bar{\rho}\right)\,\mathrm{d}\mathbf{x}\right\vert \leq \int_\Omega \left\vert\left(e^\eta\right)^2\nabla\cdot\left(\mathbf{f}*\bar{\rho}\right)\right\vert\,\mathrm{d}\mathbf{x} \\= \left\Vert e^\eta e^\eta \nabla\cdot\left(\mathbf{f}*\bar{\rho}\right)\right\Vert_1\leq \Vert e^\eta \Vert_2^2 \Vert \nabla\cdot\left(\mathbf{f}*\bar{\rho}\right)\Vert_\infty \\\leq \Vert e^\eta \Vert_2^2 \sum_{i=1}^d \Vert (f_i*\bar{\rho}_{x_i})\Vert_\infty\leq\Vert e^\eta \Vert_2^2 \sum_{i=1}^d \Vert f_i\Vert_2 \Vert\bar{\rho}_{x_i}\Vert_2,
\end{multline}
where we used H$\ddot{\mathrm{o}}$lder's, Minkowsky's, and Young's inequalities. Likewise, for the last {two} terms {of \eqref{eq:Vt_hd}},
{
\begin{multline}
    \left\vert \exp(-Kt)\int_\Omega (e^\eta)^2\nabla\cdot\mathbf{v}^0\,\mathrm{d}\mathbf{x}\right\vert \\\leq\exp(-Kt) \Vert e^\eta \Vert_2^2\sum_{i=1}^d \Vert v_{i, x_i}^0 \Vert_\infty
\end{multline}
}
\vspace{-15pt}
\begin{multline}
    \left\vert -2\mathrm{exp}(-Kt) \int_\Omega e^\eta \tilde{\mathbf{v}} \,\mathrm{d}\mathbf{x}\right\vert \leq 2\mathrm{exp}(-Kt)\Vert e^\eta \tilde{\mathbf{v}}\Vert_1 \\\leq2\mathrm{exp}(-Kt)\Vert e^\eta\Vert_2 \Vert\tilde{\mathbf{v}}\Vert_2
\end{multline}

This leads us to the following bound on the time derivative of the Lyapunov functional 
\begin{multline}
V_t \leq \left(-2D + \sum_{i=1}^d \Vert f_i\Vert_2 \Vert\bar{\rho}_{x_i}\Vert_2 \right)V\\{ +\exp(-Kt)\sum_{i=1}^d \Vert v^0_{i, x_i}\Vert_\infty V}+2\mathrm{exp}(-Kt) \Vert \tilde{\mathbf{v}}\Vert_2 \sqrt{V}. 
\end{multline}
{which is convergent due to Lemma 4 in \cite{maffettone2024leader}.\qed}
\end{pf}


\subsection{Numerical validation}
We consider a 2D mono-modal regulation scenario with a bivariate von Mises distribution (zero means, unit concentration). Agents interact through a 2D periodic Morse kernel ($L_a=\pi$, $L_r = \pi/4$, $\alpha=3.2$) with $D=0.05$, $\Phi^F = 0.2$, $M^L(0) = M^F(0) = 0.4$, $K=1$, $K_{FL}=1$, $K_{LF}=2$. Results in Fig. \ref{fig:reacting_monomodal_2d} for $t_\mathrm{f} = 100$ are qualitatively
comparable to those of 1D simulations in Figs~\ref{fig:reacting_bimodal}. Convergence of $\rho$ to $\bar{\rho}$ occurs in ~5 time units while convergence of $\rho^L$, $\rho^F$ and $\eta^F$ to their steady-state profiles is slower. We obtain final masses $M^L(t_\mathrm{f}) \approx 0.26$ and $M^F(t_\mathrm{f})  \approx 0.53$, consistent with the predicted $1/2$ ratio in Corollary  \ref{rem:mass_ratio}.

\section{Conclusions}\label{sec:conclusions}
We presented a bio-inspired leader-follower technique for spatial organization of large swarms via density control. Our strategy incorporates two crucial characteristics of natural systems: setting control objectives for the entire collective and introducing behavioral plasticity to tune the leaders-to-follower mass ratio. We derived conditions for existence and {global} stability of desired solutions, with numerical findings supporting the effectiveness, robustness, and versatility of our approach. {Numerical simulations highlights that our sufficient conditions for stability are conservative.}

Our work does not come without limitations. Analytical convergence guarantees hold at the PDE scale (infinite swarms), and convergence from continuum to discrete agent-based implementations requires assessment. Theoretical results yield exact predictions only for steady-state mass ratios; numerical simulations suggest monotonic trends, but we cannot exclude cases with higher transient role-switching than predicted. Exploring these transient dynamics may inform energetic costs of behavioral plasticity. {Furthermore, our control framework exploits full knowledge of the swarm parameters and dynamics, pointing at the necessity of a more robust architecture.} Future work should address these limitations and extend continuum descriptions beyond spatial organization to complex tasks like self-assembly and collaborative manipulation \cite{bayindir2016review}.

\begin{ack}                               
This work was supported in part by Italian Ministry of University and Research (MUR) performing the Activities of the
Project PRIN 2022 MENTOR and in part by the  NSF Grants CMMI-1932187 and EF-2222418.  
\end{ack}

\bibliographystyle{ieeetr}        
\bibliography{autosam}           



\appendix
                                        
\section{Non-plastic followers at steady-state}\label{qppE:ss_solution}
\begin{thm}
    If $\rho(x, t) = \bar{\rho}(x)$, \eqref{eq:ch6_non_react_followers} admits only the steady-state solution
    \begin{equation}\label{eqappF_etabar}
        \bar{\eta}^F(x) = \frac{\Phi^F}{\int_\mathcal{S}h(x)\,\mathrm{d}x} h(x),
    \end{equation}
    with $h$ defined as in \eqref{eqch6:h}.
\end{thm}
\begin{pf}
    Substituting $\eta^F_t = 0$, $\eta^F(x, t) = \bar{\eta}^F(x)$, and $\rho(x, t) = \bar{\rho}(x)$ into \eqref{eq:ch6_non_react_followers} leads us to \eqref{eqch6:etaF_ss}.
    Integrating in space and isolating $\bar{\eta}^F_x$ at first member, we find
    \begin{align}\label{eqappF:after_one_int}
        \bar{\eta}^F_x(x) = \frac{\bar{\eta}^F(x)(f*\bar{\rho})(x)}{D} + A,
    \end{align}
    where $A$ is an integration constant. The solution of \eqref{eqappF:after_one_int} can be written as 
    \begin{multline}\label{eqappF:sol_withA}
        \bar{\eta}^F(x) = B \,\mathrm{exp}\left[\frac{1}{D}\int (f*\bar{\rho})\,\mathrm{d}x\right] \\+ A\,\mathrm{exp}\left[\frac{1}{D}\int (f*\bar{\rho})\,\mathrm{d}x\right] \int\mathrm{exp}\left[{-}\frac{1}{D}\int (f*\bar{\rho})\,\mathrm{d}y\right]\,\mathrm{d}x,
    \end{multline}
    where $B$ is an integration constant.
    $\bar{\eta}^F$ must be periodic. The first term on the right-hand-side of \eqref{eqappF:sol_withA} is positive and periodic (it is the exponential of a periodic function). Notice that $f*\bar{\rho}$ is periodic, as a result of the circular convolution, and it sums to 0 when integrated over $\mathcal{S}$ due to Fubini's theorem for convolutions. Hence, the integral of $f*\bar{\rho}$ is itself periodic. 
    The second term on the right-hand-side of \eqref{eqappF:sol_withA}, cannot be periodic unless $A=0$, since $\mathrm{exp}\left\{{-}\frac{1}{D}\int (f*\bar{\rho})(y)\,\mathrm{d}y\right\}$
    is periodic, but it cannot sum to 0, being an exponential. Thus, $A=0$.
    For \eqref{eqappF:sol_withA} to integrate to the non-plastic followers' mass $\Phi^F$, we must have $B=\Phi_F / \int_\mathcal{S} h(x)\,\mathrm{d}x$, yielding the claim.\qed
\end{pf}

\begin{rem}
    \eqref{eqappF_etabar} is positive by construction, as $h$ is an exponential (see \eqref{eqch6:h}) and
    $\frac{\Phi_F}{\int_\mathcal{S} h(x)\,\mathrm{d}x} > 0
    $
    by construction.
\end{rem}

\end{document}